\documentclass[a4paper,11pt]{article}
\usepackage{jheppub} % for details on the use of the package, please
\usepackage{lmodern}
\usepackage{tikz}
\usetikzlibrary{decorations.pathmorphing}
\tikzset{snake it/.style={decorate, decoration=snake}}

\usepackage{slashed}
\usepackage[]{mdframed}
\usepackage[T1]{fontenc} % if needed

\newcommand{\be}{\begin{equation}}
\newcommand{\bea}{\begin{eqnarray}}

\newcommand{\ee}{\end{equation}}
\newcommand{\eea}{\end{eqnarray}}

\title{\boldmath Krylov Complexity, %conformality, 
Confinement and Universality
}
\author[a]{Ali Fatemiabhari }
\author[a]{ and Carlos Nunez}
\affiliation[a]{Centre for Quantum Fields and Gravity, Department of Physics, Swansea University, Swansea SA2 8PP, United Kingdom}
\abstract{ We perform a systematic holographic study of Krylov complexity for a wide class of confining quantum field theories. Using the geometric prescription that identifies the time derivative of the complexity with the proper momentum of a massive probe, we analyse radial geodesics in several top–down gravity duals exhibiting confinement and a mass gap.
In all geometries with a smooth infrared end–of–space we uncover a robust and universal qualitative feature: Krylov complexity exhibits oscillatory behaviour. The oscillation frequency is controlled by the confinement scale, while the amplitude depends on both the ultraviolet cutoff and the infrared scale. Additional conserved charges modify these patterns without altering their qualitative structure.
We further compare our results with the Krylov complexity of the  longitudinally perturbed Ising model. The qualitative agreement suggests that oscillatory behaviour of Krylov complexity constitutes a universal signature of confinement and provides a sensitive probe of infrared reorganisation in strongly coupled quantum field theories.

}
\begin{document} 
\maketitle
\flushbottom
%\newpage
%%%%%%%%%%%%%%%%%%%%%%%%%%%%%%%%%%%%%%%%%%%%%%%%%%%%%%%
\section{Introduction and general idea}
Quantum complexity quantifies the minimal resource cost to prepare a quantum state or implement a unitary given constrained elementary operations. Complexity has found broad application in quantum many-body systems, quantum field theory (QFT), and quantum gravity via holographic duality. {\it Nielsen} geometric complexity, {\it Krylov/spread} complexity, and tensor-network/path integral optimisation provide complementary notions for formalising complexity across these domains, for reviews see \cite{Nandy:2024evd, Baiguera:2025dkc, Rabinovici:2025otw}.

In QFT and many-body systems,  Krylov complexity is defined via the Lanczos algorithm applied to operator dynamics. A Krylov basis is constructed by successive commutators with the Hamiltonian, and the resulting operator spread encodes a complexity measure. Krylov complexity has emerged as a robust measure of operator growth and chaos, with well-defined theorems on its generic features. For a review see \cite{Rabinovici:2025otw}.
Also, Krylov complexity serves as a sensitive probe of chaotic versus integrable dynamics (see for example \cite{Baggioli:2024wbz}), and can be geometrised within dual gravitational frameworks \cite{Caputa:2021sib, Caputa:2024sux, Fan:2024iop, He:2024pox}. Present open problems include clarifying its precise holographic dual and connecting Lanczos coefficients to geometric observables in higher-dimensional AdS/CFT, see \cite{Baiguera:2025dkc,Rabinovici:2025otw} for recent reviews. In  this direction, Ref. \cite{Fatemiabhari:2025cyy} proposes a holographic Krylov complexity calculation in the dual of 4d $\cal N$=4 SYM theory, which matches with the CFT side calculation and in \cite{Fatemiabhari:2025poq}, the complexity is calculated holographically in an infinite family of 4d gauge theories, to be compared with field theory outcomes.

The examples in this work focus on complexity for the case of confining field theories. We use the holographic dual to the QFT under study. We typically start from a {\it conformal field theory}, that undergoes a deformation (via VEVs, relevant or quasi-marginal operators) and flows to a {\it confining field theory}. 
In confining gauge theories, traditional diagnostics (of confinement) rely on Wilson loops,
spectral gaps, or center symmetry. In this sense, complexity provides a novel dynamical and
information-theoretic perspective.
Below we briefly comment on papers that  connect confinement (or confinement/deconfinement
transitions) with   holographic complexity and/or
Krylov (operator spread) complexity.

On the side of the complexity-equals-volume (CV) or complexity-equals-action (CA) proposals, various papers have been written. Even when this is not the focus of our work, we mention \cite{Reynolds:2017jfs, Yang:2023qxx} that present studies on holographic solitons and analyse the CV and CA proposals. The papers \cite{Frey:2023qdv, Frey:2024tnn} study the CV proposal in the Klebanov-Strassler model (we discuss its Krylov complexity in this paper). The works \cite{Fatemiabhari:2024aua, Chatzis:2025dnu} present and develop a way to study the CV 
proposal in generic top-down holographic duals. Various studies of CV and CA in bottom-up models exist in the bibliography, we do not discuss them here.

Moving now to the Krylov complexity and its calculation in confining field theories. Interesting comments are made in Section 5.2 of \cite{Nandy:2024evd}, about the impact of an IR scale on the Lanczos coefficients. It is found there that the presence of a mass gap (or IR scale) leads to a staggering of these coefficients that change the usual behaviour of the complexity.  We are aware of other interesting papers in the bibliography in which the connection confinement-complexity is discussed. One message that follows from these works is that the complexity (an information-theory quantity) seems to detect the reorganisation of degrees of freedom when confinement takes place (and not merely in the presence of a mass gap). The operator spreading is  suppressed near the confinement scale, either growing slowly or exhibiting oscillatory behaviour (consistently with a discrete spectrum). In other words, confinement reorganises the Hilbert space in a way that complexity is sharply detecting (perhaps better than the way the entanglement entropy does). 
The papers from which we draw these lessons are \cite{Anegawa:2024wov, Fatemiabhari:2025usn, Kormos_2016, Jiang:2025wpj}. The present work studies this picture in more detail, increasing the number of confining field theories and calculating the  Krylov complexity  using the holographic dual.

\subsection{General idea of this paper}
We consider different holographic duals to confining field theories, which might (or might not) arise as deformations of a fixed point.
We use the prescription of \cite{Caputa:2024sux, Fan:2024iop, He:2024pox} to calculate the Krylov complexity. We study a massive particle, radially falling in the background in question and calculate the 'proper momentum'. We identify this geometric quantity with the time derivative of the Krylov complexity of the dual QFT. We show that the proper momentum is well defined all along the geodesic, in particular being finite.

We also study the case in which the massive particle is characterised by other quantum numbers (angular momentum or R-charge). We extend our study to various top-down holographic duals to four dimensional field theories. The various examples discussed here show similar characteristics, which lead us to propose a form of  {\it universal behaviour} of the (holographic) Krylov complexity in the case of confining field theories. The complexity presents an oscillating behaviour, with frequency and amplitude of the oscillations related to the parameters of the system, in particular, the parameter setting the scale of confinement ($\Lambda_{YM}$).

The contents of this paper are distributed as follows:
in Section \ref{section1-AR}, we study a holographic model by Anabal\'on and Ross, dual to a 3d SCFT that under a VEV deformation flows to a 2d confining and gapped QFT. We discuss analytically a simple configuration (radially falling geoedesic) and  a more elaborated one (a radial geodesic with angular momentum and R-charge). The presence of the angular momentum/R-charge introduces a novel variation to the existent studies and we analyse the qualitative changes it produces. The proper momentum is carefully calculated and the spread complexity is found using the proposal of \cite{Caputa:2024sux}. 
We discuss different qualitative aspects of the complexity obtained and draw a 
comparison with the Krylov complexity computed for a variation of the Ising model \cite{Jiang:2025wpj, Kormos_2016}.
\\
In Section \ref{section3-othermodels}, we discuss  other holographic duals to confining field theories. We present general formulas applicable to all gravity duals in the market (including conformal field theories) and then specialise these expressions for the case of models of wrapped branes, D-brane on conifolds in its various forms, etc. The picture advocated above (the oscillatory behaviour of the Krylov complexity) emerges in all these cases, with interesting subtleties that are discussed as they appear. We close with conclusions and future directions in Section \ref{concl}. Some appendices complement the presentation. The appendices discuss analytic expressions for the solution in Section \ref{sec:2.3}, comment about the definition of proper momentum and the study of complexity for a  SCFT coupled to gravity.

\section{Anabal\'on-Ross confining model}\label{section1-AR}
In this section we work with an eleven dimensional supergravity solution, presented by Anabal\'on and Ross in \cite{Anabalon:2021tua}. This was further elaborated in \cite{Anabalon:2022aig}. Various papers have studied different field theoretical aspects of the Anabal\'on-Ross class of models (among those the CV/CA proposals). Among these papers we encounter \cite{Anabalon:2022aig, Anabalon:2023lnk,Anabalon:2024che, Anabalon:2024qhf, Nunez:2023nnl, Nunez:2023nnl, Chatzis:2024kdu, Chatzis:2024top, Chatzis:2025dnu, Chatzis:2025hek, Castellani:2024ial}. Indeed, the Anabal\'on--Ross backgrounds have also been studied using the complexity-equals-volume and complexity-equals-action prescriptions. Those observables and the proper-momentum observable considered here should be regarded as complementary rather than identical probes of the same geometry. The CV/CA constructions are extended geometric observables, built from volumes or on-shell actions of bulk regions, and are naturally interpreted as state-complexity diagnostics of the full boundary theory. The present calculation instead follows the proposal of Refs.~\cite{Caputa:2024sux, Fan:2024iop, He:2024pox}, in which the rate of Krylov/spread complexity associated with a heavy and localised excitation is related to the proper momentum of a falling probe. In a confining geometry the probe motion is bounded by the UV cutoff and by the smooth end of space; this is the direct geometric origin of the oscillations found below. Thus CV/CA detect the presence and global structure of the IR cap through the renormalised bulk volume/action, while the Krylov observable studied here, which is time-dependent, detects the same IR scale dynamically through the periodic return of a probe excitation. We do not expect a numerical equality between these notions of complexity, but we do expect them to be sensitive to related IR data of the confining background.

The eleven dimensional metric and four-form \cite{Anabalon:2021tua} read,
\begin{eqnarray}
   & & ds_{11}^2= ds_4^2 + 4l^2 \sum_{i=1}^4 (d\mu_i)^2 +\mu_i^2 \left(d\varphi_i-\frac{A_1}{4l}\right)^2,\label{metric11}\\
   & & F_4= -\frac{3}{l}\text{Vol}_4 -2 l^2 \sum_{i=1}^4 \mu_i d\mu_i\wedge \left(d\varphi_i-\frac{A_1}{4l}\right)\wedge *_4 dA_1.\nonumber
   \end{eqnarray}
   
Where
   \begin{eqnarray}
   & & ds_4^2 =\frac{r^2}{l^2}\left( -dt^2+dx^2+ f(r) d\phi^2\right)+ \frac{l^2 dr^2}{r^2 f(r)},~~~f(r)= 1-\frac{\mu l^2}{r^3} -\frac{Q^2 l^2}{r^4},\nonumber\\
   & &A_1= A(r) d\phi=2Q\left(\frac{1}{r}-\frac{1}{r_*} \right) d\phi, ~~~L_\phi=\frac{4\pi l^2 r_*^3}{3r_*^4+ Q^2 l^2},\nonumber\\
   & &\mu_1=\sin\theta,~~\mu_2=\cos\theta\sin\alpha,~~\mu_3=\cos\theta\cos\alpha\sin\psi,~\mu_4= \cos\theta\cos\alpha\cos\psi.\nonumber
\end{eqnarray}
The angles range in the intervals $0\leq\varphi_i\leq 2\pi$, $0\leq \theta\leq \pi$, $0\leq \alpha\leq 2\pi$ and $0\leq\psi\leq 2\pi$. The angles $\varphi_i$, parametrise the four $U(1)$ angles of the maximal torus of the $SO(8)_R$ isometry of the internal $S^7$. Equivalently, if the round $S^7$ is embedded in $\mathbb C^4$ as $Z_i=\mu_i e^{i\phi_i}$ with $\sum_i\mu_i^2=1$, then the $\phi_i$ are the phases of the four complex coordinates. 
Aside from these, the quantities $l, \mu, Q$ are parameters characterising the solution. The parameter $\mu \neq 0$ implies that SUSY is broken (this is not the focus of this work).
To have smooth solutions, we must impose a certain period for the $\phi-$direction, given by the quantity $L_\phi$. The value of $r_*$ is the largest root of $f(r_*)=0$.

The dual field theory interpretation of this solution is similar to the one discussed in detail in the papers \cite{Anabalon:2021tua, Kumar:2024pcz, Castellani:2024ial}. Very briefly, it goes like this: 
in the far UV
we encounter the SCFT on a stack of M2 branes, this is ${\cal N}=8$ three dimensional SCFT, or the ABJM theory, a quiver $U(N)\times U(N)$ with Chern-Simons leves $k=\pm 1$. These M2 branes are compactified (with a twist) on a fixed size circle (represented by the $\phi$-coordinate in the metric above). The one form A$_1=A(r) d\phi$ performs the twist, allowing SUSY to be partially preserved (mixing the $U(1)_\phi$ with a $U(1)_{\varphi_i}^4$ inside the $SO(8)_R$ R-symmetry). As we lower the energy, KK-modes on the circle $U(1)_\phi$ decouple and we end with a two-dimensional QFT that confines. The UV-deformation as studied in \cite{Anabalon:2021tua} is mediated by a VEV for an operator of dimension three, dual to the gauge field $A_1$.

\subsection{Study of the Krylov complexity}
To study the spread complexity of this confining field theory we follow the treatment of \cite{Caputa:2024sux, Fatemiabhari:2025usn}. We probe the system with a  massive particle of mass $m$, whose trajectory is parametrised by the $t$-coordinate. The trajectory is described by the function
$r(t)$. The case of other coordinates (like $\phi,\varphi_i$, etc) being excited is  analysed  in Section \ref{sec:2.3}.
The induced metric for this particle is,
\begin{equation}
 ds_{ind}^2= dt^2\left[ -\frac{r^2}{l^2}+\frac{l^2}{r^2f(r)}\dot{r}^2 
 \right].\label{inducedmetric}
\end{equation}
The action for this particle is
\begin{eqnarray}
 & & S=\int dt L=-m \int dt \sqrt{\frac{r^2}{l^2}  -S(r)\dot{r}^2   },
 ~
 ~~\text{with}
 %~~G(r)= \frac{r^2 f(r)}{l^2} + Q^2\left(\frac{1}{r}-\frac{1}{r_*}\right)^2,~\text{and}
 ~S(r)=\frac{l^2}{r^2 f(r)}.\label{actionparticle}
\end{eqnarray}
The equation of motion is,
\begin{eqnarray}
& & -\frac{d}{dt} \left[\frac{S \dot{r}}{L} \right]=\frac{1}{2L}\left( \frac{2r(t)}{l^2} - \dot{r}^2 S' 
%- \dot{\phi}^2 G'
\right).\label{eqr}
%& & \frac{d}{dt} \left[\frac{G \dot{\phi}}{L} \right]=0.\label{eqphi}
\end{eqnarray}
\\
The system has a conserved 
%$J= \frac{\partial L}{\partial\dot{\phi}}=\frac{G\dot{\phi}}{L}$ and the 
Hamiltonian $\widehat{H}=P_r \dot{r} -L= -\frac{m^2 r(t)^2}{l^2 L}$. Using these, we  find,
\begin{equation}
%\dot{\phi}=   \frac{J}{l} \sqrt{\frac{r^2- l^2 S \dot{r}^2}{G(J^2+G)}}, ~~
\widehat{H}=m H= \frac{m r^2}{l \sqrt{(r^2- l^2 S \dot{r}^2)}}.\label{conserved}
\end{equation}
From eq.(\ref{conserved}), we obtain an expression for $\dot{r}$
\begin{equation}
 \dot{r}= \pm \frac{r(t)}{H l^2}\sqrt{\frac{H^2 l^2  - r^2}{ S(r)}} .\label{rp}
\end{equation}
Using the first order eqs.(\ref{conserved})-(\ref{rp}) and finding second derivatives expressed in terms of the variables and their first derivatives, one can check that the equation of motion (\ref{eqr}) is satisfied.

We continue solving the equation (\ref{rp}), writing
\begin{equation} \label{eq:ARr}
 \frac{t-t_0}{Hl^2}=\int_{r_{UV}}^{r} \frac{dr}{r} \times \sqrt{\frac{S(r)}{H^2 l^2 -r^2}}.    
\end{equation}
The integral runs from $r_{UV}$ to $ r(t)$, being $r(t=0)=r_{UV}$ the (large, UV) position from which we launch the massive particle. We choose the initial velocity to be zero $\dot{r}(t=0)=0$. 

In the following, we solve this equation for the supersymmetric situation ($\mu=0$). This mirrors the treatment in \cite{Fatemiabhari:2025usn} and can be solved in a  fully analytic way. After doing this, we focus on the general situation.
\subsection{The  SUSY case $\mu=0$ and its complexity} \label{sec:ARQ}
In this section, we consider the SUSY preserving situation $\mu=0$.  The function $f(r)$ is given by $f(r)=1- \frac{Q^2l^2}{r^4}$. The space ends at $r_*=\sqrt{Ql}$. We launch the massive probe from $r_{UV}$, with zero initial velocity. From the expression for the Hamiltonian (\ref{conserved}) we find $r_{UV}=l H$.
\\
To streamline the calculation, it is convenient to work in a  different coordinate $\frac{r}{l}= \frac{l}{z}$. In this coordinate, the space ranges between 
$[z_{UV}, z_*]=[\frac{l}{H},\sqrt{\frac{l^3}{Q}}]$.
%Furthermore, we set $\bar \alpha=\mathcal{H}^2$, $\beta = \mu $, $\gamma= -\mu \mathcal{H}^2$, $\zeta = 0$ and  $l=1$.
%
The integral in eq.\eqref{eq:ARr} reads (in the $z$ coordinate and setting the integration constant $t_0=0$),
\begin{align}
    &\frac{t}{H l^2} =\int_{\frac{l}{H}}^z dz~ \frac{z ~l}{{\sqrt{(H^2 z^2-l^2) \left(l^6-Q^2 z^4\right)}}}.\label{xxc}
\end{align}
In the variable $u=\frac{Q}{l^3} z^2$ and introducing the parameter $\alpha\equiv\frac{ H^2 l}{Q} $, the integral in eq.(\ref{xxc}) reads
\begin{align}
\label{eq:uQ}
    &\frac{t}{Hl^2} = \frac{1}{2 Q\sqrt{\alpha}}\int_{\frac{1}{\alpha}}^{\frac{Qz^2}{l^3}}  du \frac{1}{\sqrt{(u-1/\alpha)(1-u^2)}}.
\end{align}
The integral in eq.\eqref{eq:uQ} is in the form of an incomplete elliptic integral of the first kind. 
Let us introduce terminology and abbreviations for  elliptic functions and integrals \cite{Byrd:1971bey}. 

\begin{figure}
    \centering
    \includegraphics[width=0.7\linewidth]{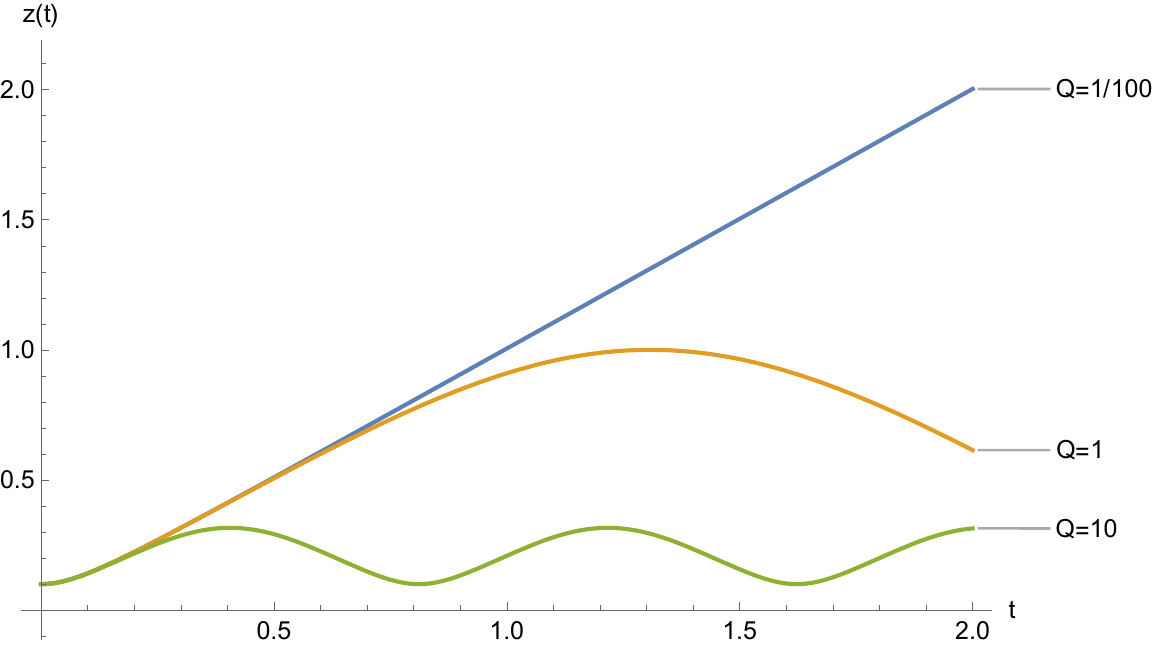}
        \includegraphics[width=0.7\linewidth]{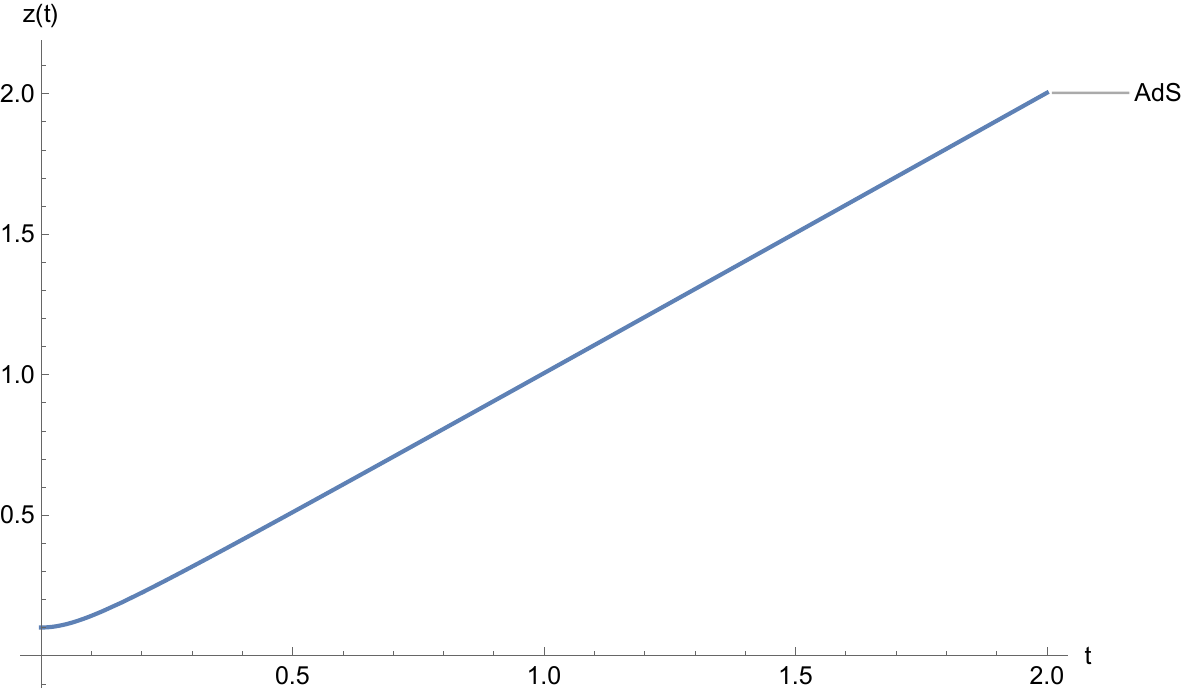}
    \caption{$z(t)$ trajectory for various values of $Q$ and comparison with the AdS case.}
    \label{fig:ztQ}
\end{figure}

% \noindent\fbox{%
%     \parbox{\textwidth}{%
\begin{mdframed}
For an integral of the form 
\begin{equation}
    \int_b^y \frac{d u}{\sqrt{(a-u)(u-b)(u-c)}},
\end{equation}
with $a, b,c$ real, $a \geq y>b, c$ one finds

\begin{align}
& \operatorname{sn}^2 \bar u=\frac{(a-c)(u-b)}{(a-b)(u-c)}, \quad k^2=\frac{a-b}{a-c}, \quad g=\frac{2}{\sqrt{a-c}} \\
& \varphi=\operatorname{am} u_1=\sin ^{-1} \sqrt{\frac{(a-c)(y-b)}{(a-b)(y-c)}}, \quad \operatorname{sn} u_1=\sin \varphi .
\end{align}
Then, 
\begin{align}
\int_b^y \frac{d u}{\sqrt{(a-u)(u-b)(u-c)}}=g \int_0^{u_1} d \bar u=g u_1 =& g \operatorname{sn}^{-1}(\sin \varphi, k) \nonumber\\
 =&g F(\varphi, k) .\label{xxxcc}
\end{align}
Here, `$\operatorname{am}$' and `$\operatorname{sn}$' are the Jacobi amplitude and elliptic sine, respectively. `$F$' is the incomplete elliptic integral of the first kind.
% }%
% }
\end{mdframed}
Comparing with the integral in eq.(\ref{eq:uQ}) we have
\begin{align}
&a=1, \quad b= \frac{1}{\alpha}, \quad c=-1,\quad g=\sqrt{2},~~~ k^2=\frac{1-\frac{1}{\alpha}}{2}, ~~y=\frac{Q}{l^3}z^2,\nonumber\\
& \varphi=\operatorname{am} u_1=\sin ^{-1}\left[\sqrt{\frac{2(y-1/\alpha)}{(1-1/\alpha)(y+1)}}\right]=\arcsin\Bigg[\sqrt{\frac{2(\alpha y-1)}{(\alpha-1)(y+1)}}\Bigg] .
\end{align}
Then 
\begin{align} \label{eq:EllipticFQ}
    &\frac{t}{Hl^2} = \frac{1}{2 Q\sqrt{\alpha}}\int_{1/\alpha}^{Qz^2/l^3} du \frac{1}{\sqrt{(u-1/\alpha)(1-u)(u+1)}}  =\frac{g}{2 Q\sqrt{\alpha}} u_1  \nonumber\\
 &  =\frac{g}{2 Q\sqrt{\alpha}} \operatorname{sn}^{-1}(\sin \varphi, k) =\frac{g}{2 Q\sqrt{\alpha}} F(\varphi, k) \nonumber\\
 &=\frac{1}{Q\sqrt{2\alpha}}F\left(\arcsin\left[\sqrt{\frac{2(\alpha y-1)}{(\alpha-1)(y+1)}} \right], \sqrt{\frac{(\alpha-1)}{2\alpha}} \right).
\end{align}
Using the relations among elliptic integrals and Jacobi functions one can invert the above relations and solve for $z(t)$. One has 
\begin{align}
    &\varphi= \operatorname{am}(u_1,k) \rightarrow \operatorname{sn}(u_1,k)\equiv \sin \operatorname{am}(u_1,k)= \sqrt{\frac{2(Qz^2/l^3-1/\alpha)}{(1-1/\alpha)(Qz^2/l^3+1)}}\nonumber \\
    & z(t)= \sqrt{\frac{l^3}{Q}}\sqrt{\frac{2+(\alpha-1)\operatorname{sn}^2 (u_1,k)}{2\alpha-(\alpha-1)\operatorname{sn}^2(u_1,k)}}.
    %\sqrt{-\frac{2/\alpha+(1-1/\alpha)\operatorname{sn}(u_1,k)^2}{-2+(1-1/\alpha)\operatorname{sn}(u_1,k)^2}}.
\end{align}
Using eqs.(\ref{eq:uQ}) and (\ref{xxxcc}), we have $u_1=\sqrt{\frac{2Q}{l^3}}t$. With this we find $z(t)$ and the original variable $r(t)$ to be,
\begin{align}
     z(t)= \sqrt{\frac{l^3}{Q}}\sqrt{\frac{2 +(\alpha-1)\operatorname{sn}^2(\sqrt{\frac{2Q}{l^3}} t,k)}{2\alpha-(\alpha-1)\operatorname{sn}^2(\sqrt{\frac{2Q}{l^3}} t,k)}},~~
     r(t)=\sqrt{Ql}\sqrt{\frac{{2\alpha-(\alpha-1)\operatorname{sn}^2(\sqrt{\frac{2Q}{l^3}} t,k)}}{2 +(\alpha-1)\operatorname{sn}^2(\sqrt{\frac{2Q}{l^3}} t,k)}}.\label{z-de-t-final}
\end{align}
In Figure \ref{fig:ztQ}, we plot $z(t)$ for $Q=1/100, 1, 10$ and $H=10$ and compare the result with the case of pure AdS in the bottom panel of Figure \ref{fig:ztQ}. As $Q$ decreases, one recovers the pure AdS case. In fact, in the simplified case treated here ($\mu=0$), $Q$ is the parameter related to the confining scale.

We  expand $z(t)$ close to boundary (at position $z_{UV}$), $z(t=0)=z_{UV}= \frac{l}{H}$ 
\begin{align} \label{eq:z0}
    z(t)= \frac{l}{H} + \frac{H^4 l^2-Q^2}{2l^3H^3}t^2 + \cdots,
\end{align}
and at the end of space $z_*=z(t_e)=\sqrt{\frac{l^{3}}{Q}}$,
\begin{eqnarray}  \label{eq:zte}
    & & z(t)= \frac{l^{3/2}}{\sqrt Q}-\frac{3 \sqrt Q}{2l^{3/2}}\left(1-\frac{ (Q/l) }{ H^2}\right)(t-t_e)^2 + \cdots.
\end{eqnarray}
It is interesting to compute the time to reach $r=r_*=\sqrt{Ql}$. The result is
\begin{equation}
 t_e=\sqrt{\frac{l^3}{2Q}}K\left( \frac{(Q-H^2 l)^2}{4H^4l^2}\right),\label{halfperiod}  
\end{equation}
where $K$ represents Jacobi elliptic function. Interestingly, the period of the motion is inverse with the parameter $Q$ which sets the scale of confinement. 
\subsection{The proper momentum and the complexity}
The rate of change of the complexity is  defined in terms of the \textit{proper momentum} $P_{\bar{y}}$ as \cite{Caputa:2024sux}.
One may consider to define the derivative of the complexity in terms of the (usual) momentum, for example
\begin{eqnarray}
& & \dot{C}(t)\sim P_r,~~\text{or}~~\dot{C}(t)\sim P_z.\label{definitionsCdot}\\
& & \text{with}~~P_r=\frac{\partial L}{\partial \dot{r}}=\frac{m S(r) \dot{r}}{\sqrt{\frac{r^2}{l^2}- S(r) \dot{r}^2}} ~~\text{and}~~ P_z=\frac{\partial L}{\partial\dot{z}}=\frac{m l \dot{z}}{f(z) z \sqrt{1-\frac{\dot{z}}{f(z)}}}.\nonumber
\end{eqnarray}
There is nevertheless an issue with the definitions in eq.(\ref{definitionsCdot}), these expressions diverge close to the end of the space, as the function $f$ (that vanished at the end of the space) appears in the denominator. In \cite{Caputa:2024sux}, the authors made a definition, specially tailored to match a geodesic calculation in AdS$_3$ with a generic CFT calculation. One unintended virtue of this definition is that it avoids the divergent behaviour close to the end of the space. See Appendix \ref{appendixBproper} for a study of this.

In fact, the authors of \cite{Caputa:2024sux}
define a coordinate $\bar{y}$ such that when all coordinates are fixed (except for the 'radial coordinate') we find $ds^2=d\bar{y}^2=\frac{l^2}{r^2 f(r)} dr^2$. This defines the $\bar{y}$-coordinate and we calculate the proper momentum.
More explicitly, in the original $r$-coordinate or in the $z$-coordinate we find,
\begin{equation}
P_{\bar{y}}=
\frac{\partial L}{\partial\dot{r}}\times \frac{d \dot{r}}{d\dot{\bar{y} }}
= \frac{\partial L}{\partial\dot{z}}\times \frac{d\dot{z}}{d\dot{\bar{y}}}.\label{Py}
\end{equation}
Leading to a rate of change of the complexity
\begin{align}
   \partial_t C(t) &=-\frac{P_{\bar y}}{\epsilon} \nonumber\\
    P_{\bar y}&=P_{z} \frac{\partial \dot z}{\partial \dot {\bar y}}=- \frac{m\dot{z}}{l \sqrt{f(z)}}\frac{1}{\sqrt{1 -\frac{\dot{z}^2}{f(z)}}} \;.\label{derivativeofC}
\end{align}
Here, $\epsilon$ is a non-universal positive constant which is model-dependent and can be fixed while comparing with a field-theoretic calculation. It is worth mentioning that the Krylov complexity depends on the mass of the probe particle. In the case of CFT calculations in \cite{Caputa:2024sux}, this dependence implied that the Krylov complexity of the CFT is proportional to the scaling dimension of the boundary operator that induces the out-of-equilibrium dynamics.

The expansions of this quantity close to the boundary and end of space are
\begin{align} \label{eq:pbarrhoexp}
 \partial_t C(t)   \propto &P_{\bar y}\Bigg|_{t\sim0} =- \frac{m\sqrt{H^4-(Q/l)^2}}{H l} t +O(t)^{2}\;, \\
 &  P_{\bar y}\Bigg|_{t\sim t_e} = \mp\frac{m\sqrt{H^2-(Q/l)}}{\sqrt{Q/l}} +O(t-t_e)^{2}\;. \label{complexitylarget}
\end{align}

\begin{figure}
    \centering
    \includegraphics[width=0.7\linewidth]{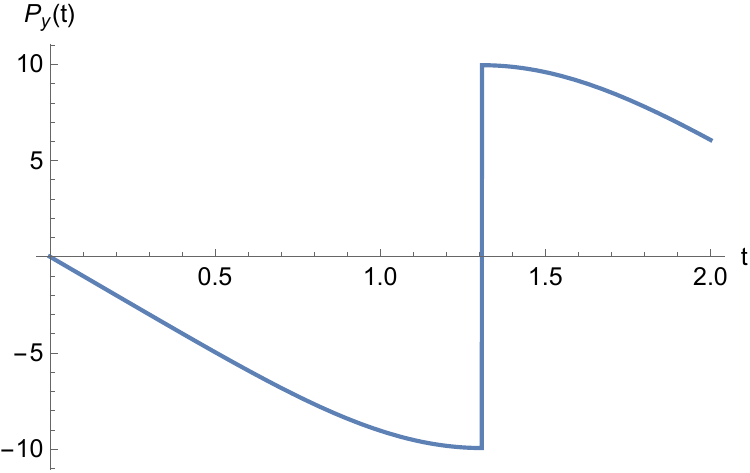}
        \includegraphics[width=0.7\linewidth]{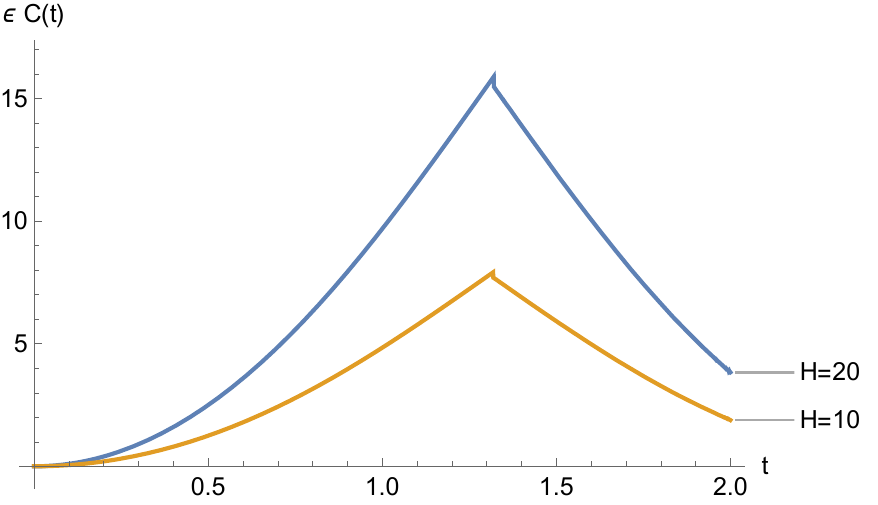}
    \caption{$P_{\bar{y}}(t)$ (upper panel) for $H=10$,  $m=1$ and $C(t)$ (lower panel) for various values of $H$, $m=1$.}
    \label{fig:PtQ}
\end{figure}

The proper momentum and complexity are depicted in Figure \ref{fig:PtQ}. We observe that the complexity has an oscillating behaviour, here found analytically in terms of Jacobi special functions. The intuition developed in \cite{Fatemiabhari:2025usn} is that we have a QFT with a UV cutoff (represented by $r_{UV}$ in the dual geometry) and an IR minimal energy (represented by $r_{*}$). This implies a system with a finite number of degrees of freedom, which is typically associated with oscillation in the complexity \cite{Baiguera:2025dkc}. The complexity grows, as the particle falls from $r_{UV}$ and decreases as the particle bounces back and climbs up from $r_*$, as suggested in \cite{Susskind:2018tei,Ageev:2018msv}.

It is worth mentioning that the jumps in the proper momentum $P_{\bar y}$ observed in Figure \ref{fig:PtQ} are  an artifact of the coordinate system used and do not lead to a discontinuity in the total momentum or energy of the particle. The particle starts its motion in $\bar y$ direction from rest, while it stays at fixed values of other coordinates $x=x_0, \phi=\phi_0, \cdots$ as a result of symmetries of the action. Close to the end of space,  $(\bar y,\phi)$ coordinates can be considered as a polar coordinate system for a flat two dimensional plane in which $\phi$ is degenerate at $\bar y =\bar y_*$ \footnote{Note that close to the end of space, one can approximate the metric in the $(\bar y,\phi)$ plane as follows
\begin{align*}
    ds^2_{(\bar y,\phi)}|_{\bar y\sim \bar y_*}\sim d \bar y^2+\frac{r(\bar y)^2}{l^2}f(\bar y) \phi^2\sim \frac{l^2 dr^2}{r^2 f(r)}+ \frac{r^2f(r)}{l^2}d\phi^2\sim d \tilde r^2 + \frac{\tilde r^2}{L_\phi^2} d \phi^2,
\end{align*}
where $\tilde r=\sqrt{r-r_*}$. The momentum of the particle, $P_{\tilde r}\propto P_{\bar y}$ moving towards $\tilde r =0$ will change sign after passing the origin. In a Cartesian coordinate defined as 
\begin{align*}
\tilde x_1= \tilde r\cos(\phi/L_\phi),\quad \tilde x_2= \tilde r\sin(\phi/L_\phi),
\end{align*}
with particle presumably moving in $\tilde x_1$ direction $(\phi(t=0)=0)$, the discontinuity will not occur.
}.
The particle reaches and passes the end of space with certain momentum but the value $P_{\bar y}$ flips sign and $\phi_0\to \phi_0+L_\phi/2$ as is usual for centripetal motion formulated in polar coordinates. This is the reason for the appearance of $\mp$ sign in Eq.\eqref{complexitylarget} before and after $t=t_e$. While the momentum changes sign, the absolute value is continuous.    

It is natural to ask what is the {\it field theory calculation} that gives the complexity in eq.(\ref{derivativeofC}). According to \cite{Nandy:2024evd, Anegawa:2024wov}, one should have the masses of particles in the theory. In our case, the masses of all glueballs. These glueballs can be considered to be non-interacting in the large $N$ limit. Then a given two point function should be computed, from which the Lanczos coefficients and Krylov complexity can be calculated. The first step is to have the spectrum of glueballs. Our models have scalar, spin-half, spin-one, spin-three-halves and spin-two glueballs. May be some subsector is enough to have a good estimate of the complexity. These glueball's masses can be calculated following the formalism developed in \cite{Bianchi:2003ug,Berg:2005pd,Berg:2006xy,Elander:2009bm,Elander:2010wd,Fatemiabhari:2024lct}. Once we have these masses, we can follow the treatment in \cite{Anegawa:2024wov}. It may be interesting to compare with the treatment in \cite{Caputa:2024xkp} and see if some similarities can be found.
\\
Let us be more precise about what a direct field-theory calculation would require. One should choose a gauge-invariant operator $\mathcal O$ whose bulk dual is the fluctuation or probe being considered and compute its real-time two-point function in the confining theory. At large $N$ this correlator has a discrete spectral representation,
\begin{equation}
 G_{\mathcal O}(t)=\sum_n \rho_n e^{-iE_n t},
 \qquad
 \rho_n=|\langle 0|\mathcal O|n\rangle|^2,
\end{equation}
where the states $|n\rangle$ are glueball states in the appropriate spin and charge channel. The moments of this spectral density determine the Lanczos coefficients and therefore the Krylov complexity. Thus the calculation requires both the glueball masses $E_n$ and the spectral residues $\rho_n$. The masses can be obtained from the standard gauge-invariant fluctuation problem in the corresponding supergravity background, but the extraction of enough residues and the subsequent Lanczos analysis is a substantial numerical project. We therefore view a quantitative boundary calculation as feasible in principle, but beyond the scope of the present paper. Of course, we would need {\it all} glueball states in order to check or rule-out the holographic result.

\subsection{Complexity in the presence of Angular Momentum}\label{sec:2.3}

In this section, we aim to let the massive particle probe also the compact $\phi$ coordinate besides the radial $r$ coordinate. It turns out that a consistent solution requires the particle to also move in the internal deformed sphere parametrized by $\varphi_i$. This is not surprising as the coordinates of the $U(1)^4$ R-symmetry are locked with the $\phi$-coordinate by the twist represented by the one form $A_1$. Hence, the induced metric by choosing $\theta=\alpha=\psi=0$ is
\begin{eqnarray}
    ds_{ind}^2= dt^2\left[ -\frac{r^2}{l^2}+\frac{l^2}{r^2f(r)}\dot{r}^2 +\frac{r^2f(r)}{l^2}\dot\phi^2+{4l^2}(\dot \varphi_4-\frac{A(r)}{4l}\dot\phi)^2
 \right].\label{inducedmetricphi}
\end{eqnarray}
hence, the trajectory is parametrized by $(r(t),\phi(t),\varphi_4(t))$ functions. 
The action for this particle is
\begin{eqnarray}
 & & S=\int dt \mathcal{L}=-m \int dt \sqrt{\frac{r^2}{l^2}  -S(r)\dot{r}^2 -G(r)\dot\phi^2-{4l^2}\left(\dot \varphi_4-\frac{A(r)}{4l}\dot\phi\right)^2  },\\
 ~
&& ~~\text{with}
 ~~G(r)= \frac{r^2 f(r)}{l^2},~\text{and}
 ~S(r)=\frac{l^2}{r^2 f(r)}. \nonumber\label{actionparticlephi}
\end{eqnarray}
The equations of motion are,
\begin{eqnarray}
& & -\frac{d}{dt} \left[\frac{S \dot{r}}{L} \right]=\frac{1}{2L}\left( \frac{2r(t)}{l^2} - \dot{r}^2 S' 
- \dot{\phi}^2 G'-\frac{A'(r)A(r)}{2}\dot\phi^2{+ 2 l}{A'(r)}\dot\phi\dot \varphi_4
\right),\label{eqrphi}\\
& & \frac{d}{dt} \left[\frac{{{G\dot{\phi}+\frac{A^2}{4}\dot{\phi}- l A\dot{\varphi}_4}}
%G \dot{\phi}-\frac{A(r)}{2l}(\dot \varphi_4-\frac{A(r)}{4l}\dot\phi)
}{L} \right]=0,\\
& & \frac{d}{dt} \left[\frac{
\dot \varphi_4-\frac{A(r)}{4l}\dot\phi
}{L} \right]=0.\label{eqphiphi}
\end{eqnarray}
where $L=\sqrt{\frac{r^2}{l^2}  -S(r)\dot{r}^2 -G(r)\dot\phi^2-{4l^2}(\dot \varphi_4-\frac{A(r)}{4l}\dot\phi)^2  }$.
\\
The system has three conserved quantities
\begin{eqnarray}
& & {\hat J=mJ}= \frac{\partial \mathcal{L}}{\partial\dot{\phi}}=m\frac{
{G\dot{\phi}+\frac{A^2}{4}\dot{\phi}- l A\dot{\varphi}_4}
%G \dot{\phi}-\frac{A(r)}{2l}(\dot \varphi_4-\frac{A(r)}{4l}\dot\phi)
}{L}, ~~~ {\hat J_4=mJ_4}= \frac{\partial \mathcal{L}}{\partial\dot{\varphi_4}}=m\frac{\dot \varphi_4-\frac{A(r)}{4l}\dot\phi}{L},\nonumber\\
& &  
\widehat{H}=mH=P_r \dot{r}+P_\phi \dot{\phi}+P_{\varphi_4} \dot{\varphi_4} -\mathcal{L}= \frac{{m}~ r(t)^2}{l^2 L}.\label{conservedquantx}
\end{eqnarray}
In what follows, we assume $J_4=0$ for simplicity, hence $\dot \varphi_4=\frac{A(r)}{4l}\dot\phi$. Using these, we  find
\begin{eqnarray}
& & \dot{\varphi}_4= \frac{J A(r) r^2}{4 H l^3 G(r)}, ~~~\dot{\phi}= \frac{J r^2}{H l^2 G(r)}, \nonumber\\
& & \dot{r}=\pm \frac{r(t)}{Hl^2}\sqrt{\frac{H^2l^2 G(r)-\left(~J^2+ G(r)~\right)r^2}{G(r) S(r)}}.\label{betterex}
\end{eqnarray}
\begin{equation}
%\dot{\phi}=   \frac{J}{l} \sqrt{\frac{r^2- l^2 S \dot{r}^2}{G(J^2+G)}}, ~~
\widehat{H}=m H= \frac{r^2\sqrt{J^2+G}}{l \sqrt{G(r^2- l^2 S \dot{r}^2)}}.\label{conservedphi}
\end{equation}
% From eq.(\ref{conservedphi}), we obtain an expression for $\dot{r}$
% \begin{equation}
%  \dot{r}= \pm \frac{r(t)}{H l^2}\sqrt{\frac{H^2 l^2 G(r)- J^2 r^2 - G(r)r^2}{G(r) S(r)}} .\label{rpphi}
% \end{equation}
Using the first-order eq.(\ref{betterex}) and finding second derivatives expressed in terms of the variables and their first derivatives, one can check that the equation of motion (\ref{eqr}) is satisfied.

We solve the equation (\ref{betterex}), writing
\begin{equation} \label{eq:ARrphi}
 \frac{(t-t_0)}{Hl^2}=\int \frac{dr}{r} \times \sqrt{\frac{G(r) S(r)}{H^2 l^2 G(r)-r^2(J^2+G(r))}}    %
 \end{equation}
In Appendix \ref{exactresulJ}, this integral is analytically studied. 

We now move to the definition of the proper momentum. Below we derive an expression in terms of the three active coordinates $[r,\phi,\varphi_4]$ that is easily generalised to more coordinates, and shows clearly the logic in defining the proper-coordinate $y$.
\subsubsection{Proper momentum}
In this section, we calculate the momentum in a new coordinate defined as
\begin{equation}
     ds^2 \equiv d\bar y^2 =\Bigg[S(r)d{r}^2+G(r)d\phi^2+{4l^2}\left(d \varphi_4-\frac{A(r)}{4l}d\phi\right)^2\Bigg].
\end{equation}
The momentum in this coordinate matches the criteria introduced in \cite{Caputa:2024sux} as the coordinate parametrizing distances naturally along the geodesic. 

Since $r(t)$, $\phi(t)$ and $\varphi_4(t)$ are functions of time, one can (in principle) invert these relations to obtain $\phi(r)$,  $\varphi_4(r)$, and similarly $r(\phi), \varphi_4(\phi)$, or $r(\varphi_4), \phi(\varphi_4)$. We have 
\begin{align}
    & d\bar y = \sqrt{S(r) +G(r)\phi'(r)^2+{4l^2}\left( \varphi_4'(r)-\frac{A(r)}{4l}\phi'(r)\right)^2} ~dr \nonumber \\
    & \rightarrow \frac{d\bar y}{dr}=\frac{\dot {\bar y}(t)}{\dot r (t)}=\sqrt{S(r) +G(r)\phi'(r)^2+{4l^2}\left( \varphi_4'(r)-\frac{A(r)}{4l}\phi'(r)\right)^2},\nonumber \\
    & d\bar y =\sqrt{S(r)r'(\phi)^2 +G(r)+{4l^2}\left( \varphi_4'(\phi)-\frac{A(r)}{4l}\right)^2}~ d\phi \nonumber \\
    & \rightarrow \frac{d\bar y}{d\phi}=\frac{\dot {\bar y}(t)}{\dot \phi (t)}=\sqrt{S(r)r'(\phi)^2 +G(r)+{4l^2}\left( \varphi_4'(\phi)-\frac{A(r)}{4l}\right)^2},\nonumber\\ 
    & d\bar y = \sqrt{S(r)r'(\varphi_4)^2 +G(r)\phi'(\varphi_4)^2+{4l^2}\left( 1-\frac{A(r)}{4l}\phi'(\varphi_4)\right)^2} ~d\varphi_4 \nonumber \\
    & \rightarrow \frac{d\bar y}{d\varphi_4}=\frac{\dot {\bar y}(t)}{\dot \varphi_4 (t)}=\sqrt{S(r)r'(\varphi_4)^2 +G(r)\phi'(\varphi_4)^2+{4l^2}\left( 1-\frac{A(r)}{4l}\phi'(\varphi_4)\right)^2}.\label{toy}
\end{align}
 Here, the derivatives with primes are with respect to the variable the function depends on. Also $\dot {\bar y}(t) =\sqrt{S(r)\dot{r(t)}^2 +G(r)\dot\phi(t)^2+{4l^2}\left(\dot \varphi_4(t)-\frac{A(r)}{4l}\dot\phi(t)\right)^2} $. One can derive the following relations
\begin{align}
    & \frac{\partial\dot r}{\partial \dot {\bar y}}=\frac{1}{\sqrt{S(r) +G(r)\phi'(r)^2+{4l^2}\left( \varphi_4'(r)-\frac{A(r)}{4l}\phi'(r)\right)^2}}\nonumber \\
    & =\frac{\dot r}{\sqrt{S(r)\dot{r}^2 +G(r)\dot\phi^2+{4l^2}\left(\dot \varphi_4-\frac{A(r)}{4l}\dot\phi\right)^2}}=\frac{\dot r(t)}{\dot {\bar y}(t)},\nonumber \\
    & \frac{\partial\dot \phi}{\partial \dot {\bar y}}=\frac{1}{\sqrt{S(r)r'(\phi)^2 +G(r)+{4l^2}\left( \varphi_4'(\phi)-\frac{A(r)}{4l}\right)^2}}=\frac{\dot \phi(t)}{\dot {\bar y}(t)},\nonumber\\ 
    & \frac{\partial\dot \varphi_4}{\partial \dot {\bar y}}=\frac{1}{\sqrt{S(r)r'(\varphi_4)^2 +G(r)\phi'(\varphi_4)^2+{4l^2}\left( 1-\frac{A(r)}{4l}\phi'(\varphi_4)\right)^2}}=\frac{\dot \varphi_4(t)}{\dot {\bar y}(t)}.\label{yderivatives}
\end{align}

% We used the notation $\eta'=\frac{d\eta}{dr}$ and $r'=\frac{dr}{d\eta}$ and that $\eta'(r)=\frac{1}{r'(\eta)}$ in the very last expression of eq.(\ref{toy}).
 We define the proper momentum in the $y$ direction as
\begin{align}
    & P_{y}=\frac{\partial\mathcal{L}}{\partial \dot {\bar y}}=\frac{\partial\mathcal{L}}{\partial \dot {r}}\frac{\partial\dot r}{\partial \dot {\bar y}}+\frac{\partial\mathcal{L}}{\partial \dot {\phi}}\frac{\partial\dot \phi}{\partial \dot {\bar y}}+\frac{\partial\mathcal{L}}{\partial \dot {\varphi_4}}\frac{\partial\dot \varphi_4}{\partial \dot {\bar y}}\label{propermomentumpy}\\
    & = P_{r}\frac{\dot r(t)}{\dot {\bar y}(t)}+P_{\phi}\frac{\dot \phi(t)}{\dot {\bar y}(t)}+P_{\varphi_4}\frac{\dot \varphi_4(t)}{\dot {\bar y}(t)}.\nonumber
\end{align}
%Recalling that
%\begin{align}
     %\eta'(r)=\frac{1}{r'(\eta)}=\frac{\dot \eta(t)}{\dot r(t)},
%\end{align}
%we calculate the momentum without necessarily inverting the expression for the trajectory of the particle.

Using the action in eq.\eqref{actionparticlephi} the  proper momentum simplifies considerably as one has
\begin{align}
  &  P_r=\frac{mS(r)\dot r(t)}{L},\\
   & P_{\phi}=\frac{m\left[G(r)\dot{\phi} + l A(r) (\dot{\varphi}_4 - \frac{A(r)\dot{\phi}}{4l}) \right]     
   %G \dot{\phi}-\frac{A(r)}{2l}(\dot \varphi_4-\frac{A(r)}{4l}\dot\phi)
   }{L},\\
& P_{\varphi_4}=4ml^2\Big(\frac{\dot \varphi_4-\frac{A(r)}{4l}\dot\phi}{L}\Big)\;,
\end{align}
which with $L={\sqrt{\frac{r^2}{l^2}  -S(r)\dot{r}^2 -G(r)\dot\phi^2-{4l^2}\left(\dot \varphi_4-\frac{A(r)}{4l}\dot\phi\right)^2  }}$ gives
\begin{align}
  &  P_{\bar y}=\frac{m}{L \dot {\bar y} (t)}\Bigg(S(r)\dot{r}^2 +G(r)\dot\phi^2+{4l^2}\left(\dot \varphi_4-\frac{A(r)}{4l}\dot\phi\right)^2 \Bigg)=\frac{m\dot {\bar y} (t)^2}{L \dot {\bar y} (t)}=\frac{m\dot {\bar y} (t)}{L}. \label{eq:ymomentum}
\end{align}

\begin{figure}
    \centering
    \includegraphics[width=0.5\linewidth]{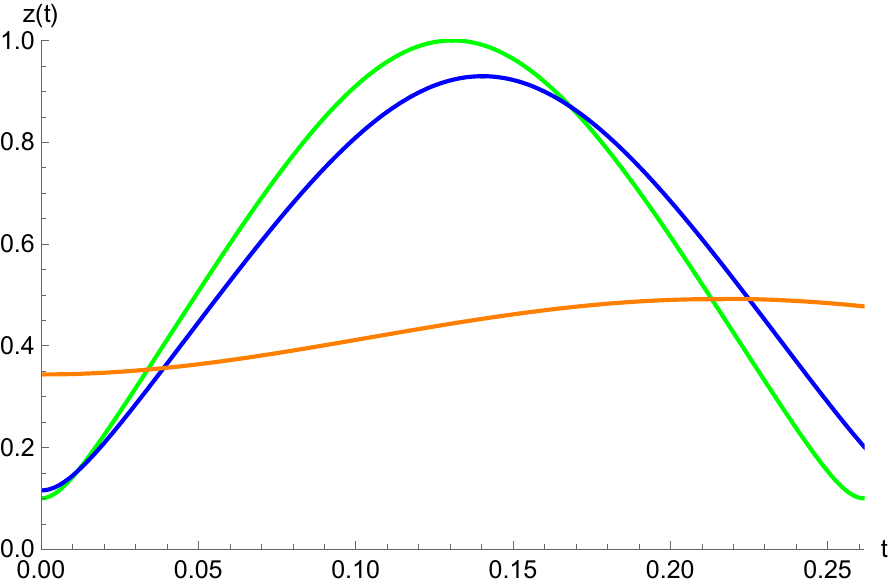}
        \includegraphics[width=0.5\linewidth]{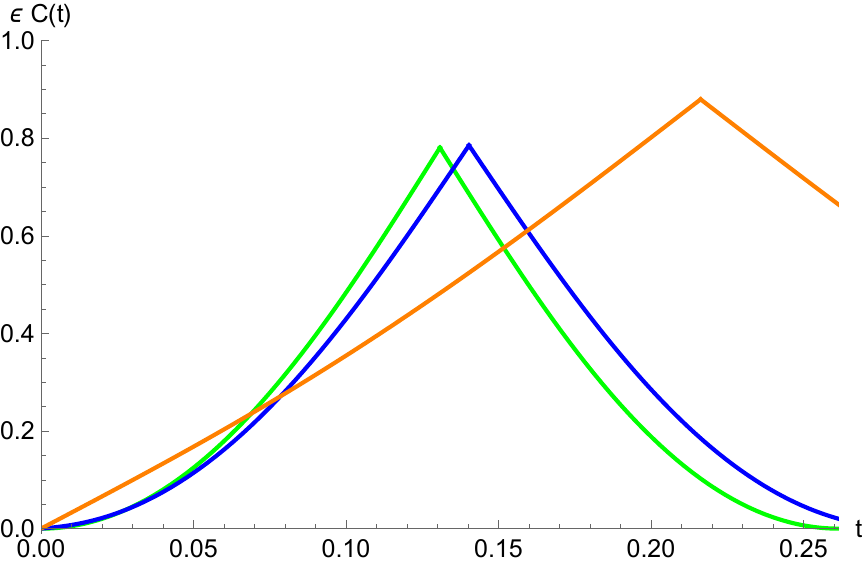}
             \includegraphics[width=0.5\linewidth]{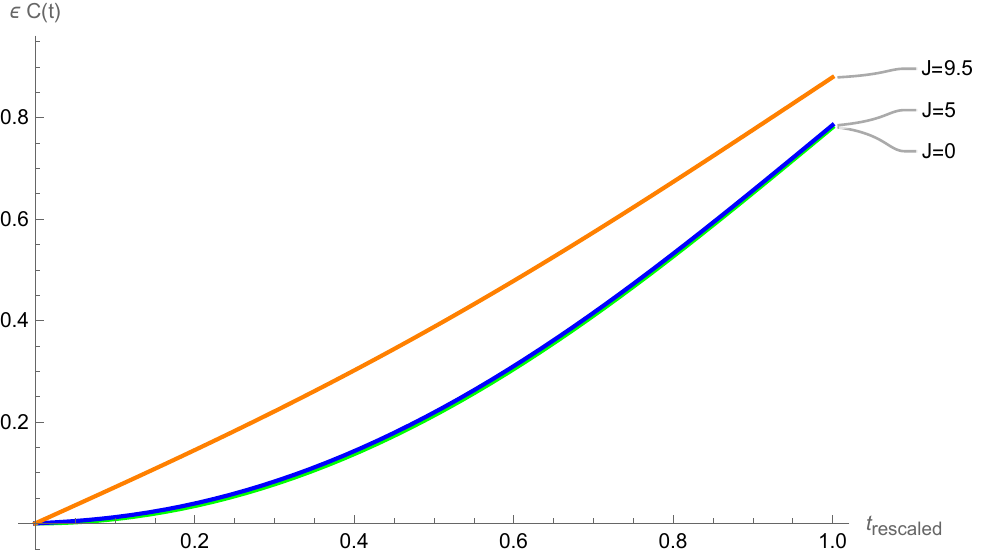}
    \caption{$r(t)$, $C(t)$ and $C(t)$ with rescaled time. The calculation results for $J=0$(Green), $J=5$(Blue), and $J=9.5$(Orange) are depicted with $H=10$. The rescaling is done such that the frequency of the oscillations matches for different values of J. $m=1$ is chosen throughout the calculation.}
    \label{fig:rtJ}
\end{figure}

\begin{figure}
    \centering
    \includegraphics[width=0.5\linewidth]{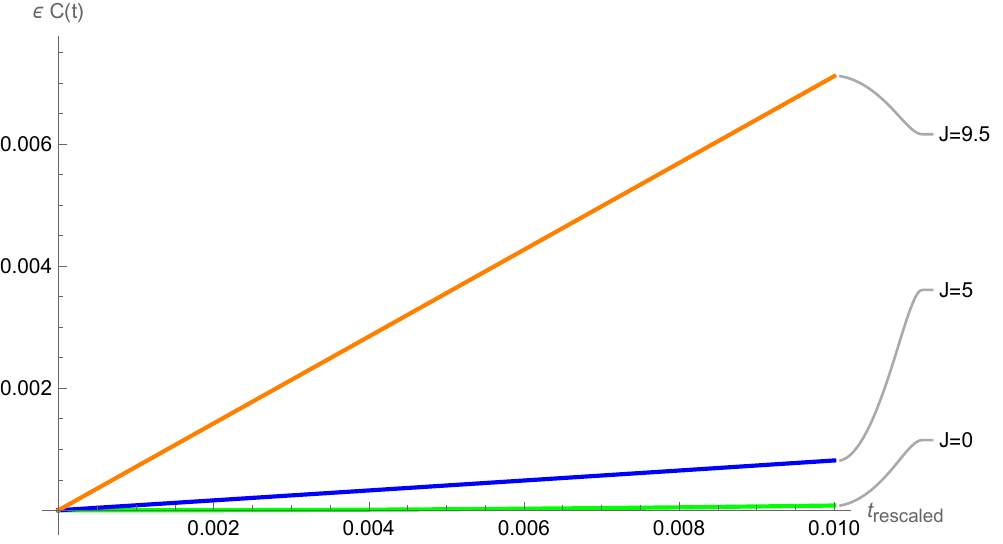}
        \includegraphics[width=0.5\linewidth]{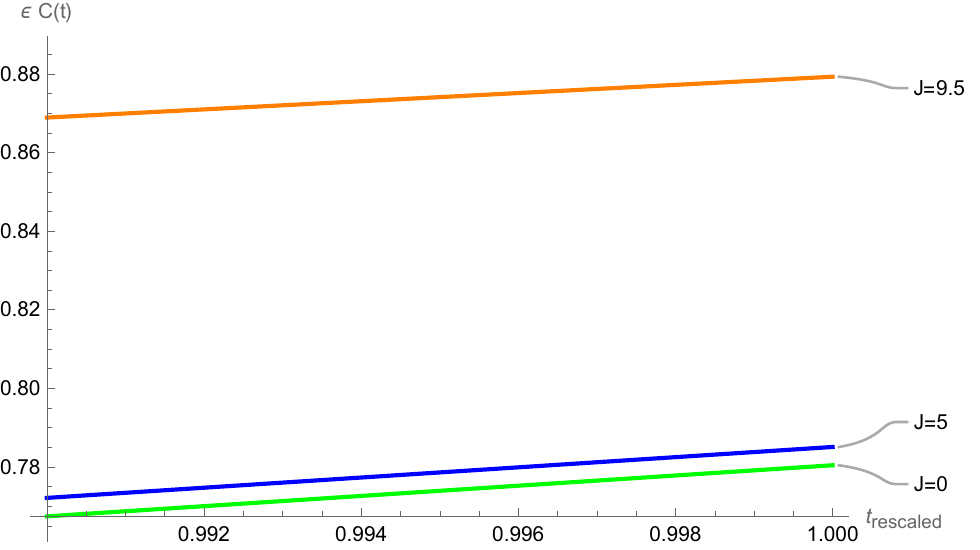}
    \caption{$C(t)$ with rescaled time  for different J values and $H=10$ in early times (upper panel) and late times (lower panel). The rescaling is done such that the frequency of the oscillations matches for different values of J. $m=1$ is chosen throughout the calculation.}
    \label{fig:rtJ2}
\end{figure}

This has a useful form. We mention a more straightforward method to derive this relation. Since there is only one free parameter along the geodesic, one can in principle solve $r(t)$ in terms of $y(t)$ as $r(y)$. Then, the action in eq.\eqref{actionparticlephi} can be written as
\begin{eqnarray}
    S=-m \int dt \sqrt{\frac{r(\bar y(t))^2}{l^2}  -\dot{\bar y}^2  }.
\end{eqnarray}
Hence from the definition $P_y=\frac{\partial\mathcal{L}}{\partial \dot {\bar y}}$ one finds
\begin{eqnarray}
    P_{\bar y}=\frac{m\dot {\bar y} (t)}{L}.
\end{eqnarray}
Inserting eq.\eqref{betterex} in this expression, one obtains
\begin{eqnarray}
    P_{\bar y}=m\sqrt{H^2/r(t)^2-1}.
\end{eqnarray}
We plot the trajectory of the particle and its complexity in Figures \ref{fig:rtJ} and \ref{fig:rtJ2}.
%%%%

From the numerical study of the solutions we learn that the presence of the conserved charge $J$ (angular momentum),
decreases both the frequency and amplitude of the oscillations of the coordinate $z(t)$. At the same time, larger values of $J$ imply larger values (amplitude) of the complexity, but smaller frequency of the complexity-oscillations. In Appendix \ref{exactresulJ}, we perform an analytic study, showing that complexity changes qualitatively for early times (growing linearly, instead of quadratically in time). At late times, the effects of $J$ are just a multiplicative factor compared to the case in eq.(\ref{complexitylarget}). The fact that the quantum number $J$ influences the early stages of the evolution but is does not make a major difference at late times is reminiscent of \cite{Caputa:2025mii, Caputa:2025ozd}. 

It is worth commenting briefly about the relation with \cite{Caputa:2025mii, Caputa:2025ozd} and the change in regime at short times. As we explain in Appendix \ref{exactresulJ}, at short times the complexity behaves linearly with time. 
%\begin{equation*}
%C(t\sim0)\Big|_{J=0}\sim t^2 ~~\text{to}~~C(t\sim0)\Big|_{J\neq0}\sim t.   
%\end{equation*}
Details provided in Appendix \ref{exactresulJ}.
\\
For $J\neq0$ the particle is released with $\dot r(0)=0$ but not with zero velocity in the full configuration space. The conserved angular momentum forces non-zero motion along the internal directions already at $t=0$. Consequently the proper momentum along the full trajectory has a non-zero initial value. The unrenormalised integral of $-\frac{P_y}{\epsilon}$ therefore contains a kinematic {\it drift} contribution. This should not be interpreted as a claim that ordinary Krylov/spread complexity has a linear initial growth. In the standard Krylov construction, with the initial state chosen as the first Krylov vector, the complexity starts quadratically. The finite-$J$ result should instead be viewed as an exploratory charged-probe observable. A precise boundary dual would require a charge-resolved Krylov construction, or an appropriate subtraction of the background charged drift, and we leave this more refined comparison for future work. In Appendix \ref{exactresulJ}, we provide a tentative way of understanding the {\it drift} term. Another possible cure for this unexpected linear behaviour, is to consider a particle 'smeared' on the compact direction, with fixed angular momentum. The smearing might produce a well defined state of fixed angular momentum on the field theory side, as compared with the one produced with a localised particle, that we studied here. See also \cite{Berenstein:2019tcs}.  This may recover the quadratic behaviour of the spread complexity of the dual state at short times. We leave this possibility for future study.
\\
On the other hand, the comparison with Refs.~\cite{Caputa:2025mii,Caputa:2025ozd} should be understood only at the {\it qualitative level}, that conserved charges refine the Krylov problem by decomposing the dynamics into charge sectors and by changing the weights with which those sectors contribute. We do not mean to suggest that the universal early-time behaviour of standard Krylov complexity is altered in those works. At large times the behaviour we find, like the one in \cite{Caputa:2025mii, Caputa:2025ozd} is universal. In our finite-$J$ geodesic the non-zero initial proper momentum is a kinematic effect of the charged classical trajectory, as explained above, and a precise comparison with the boundary theory would require a holographic prescription for the charge-resolved Krylov construction.

\subsection{Comparison with the Ising model }
%of \cite{Jiang:2025wpj, Kormos_2016}}
Let us attempt to connect our study above with that of the papers \cite{Kormos_2016,Jiang:2025wpj}. The authors of these papers study a variation of the Ising model in which a 'transversal' magnetic field is applied. The Hamiltonian reads
\begin{equation}
 H=-J\sum_{i=1}^N \sigma_i^{(z)}\sigma_{i+1}^{(z)} +h_x \sigma_i^{(x)} +h_z \sigma_i^{(z)}.\label{ising}   
\end{equation}
If the parameter $h_z=0$ the model is integrable. If also $h_x=1$ the model is a CFT in the continuum limit. For $h_x<1$ we are in the ferromagnetic phase, on which we are interested (otherwise, for $h_x>1$ we are in the paramagnetic phase).

For $h_z=0, h_x<1$, the excitations are freely moving domain walls (which separate domains of different magnetisation). If we switch on $h_z>0$, the $Z_2$-symmetry is broken and a linear potential between these domain walls is generated. The walls are confined in pairs, forming mesons \cite{Kormos_2016}.

The Krylov complexity is calculated applying the Lanczos algorithm and after a quench (a sudden variation of the $h_z$ parameter). The authors of \cite{Jiang:2025wpj}
show that well inside the ferromagnetic phase (they choose $4h_x=1$) the complexity is  large for $h_z=0$ and decreases abruptly for non-zero $h_z$. Also, oscillations appear in the  complexity for non-zero $h_z$. Qualitatively, the amplitude of these oscillations decrease, and the frequencies grow as $h_z$ grows above zero.

For our holographic model the dual 3d SCFT is twisted-circle compactified and flows to a gapped and confining QFT in two dimensions.
%--quite different from the Ising model in %eq. (\ref{ising})
On the gravity (holographic) side, this RG-flow is mediated by the functions $f(r)$ and $A(r)$. We have two scales, one set by the parameter $Q$ and related to the confinement scale (holographically the value of $r_*$). The second parameter is the UV scale introduced as a regulator $r_{UV}$ and related to the Hamiltonian $H$.  In the Ising system there are also two scales: the size of the system and $h_z$ (for fixed $h_x$).

In both systems oscillations in the complexity are observed. The frequency of the oscillations  grow as $Q$ grows--see eq.(\ref{halfperiod}) and  Figure \ref{fig:ztQ}. A natural association $Q\leftrightarrow h_z$ arises. From eq.(\ref{complexitylarget}), we find that the amplitude of the oscillations in our model scales as $A^2\sim \frac{l^2 H^2}{Q}$, showing dependence on the UV-cutoff and also inverse dependence on the confining scale (in coincidence with the Ising model that shows for fixed size $L$, a decreasing amplitude in terms of $h_z$). From Figure \ref{fig:PtQ} and eq.(\ref{complexitylarget}), we find that the larger the UV-cutoff $H$, the larger is this amplitude of the oscillation. 
\\
There is an important caveat in the analogy with the spin-chain. A finite spin chain has a finite-dimensional Hilbert space, oscillations and recurrences can occur even without confinement. Therefore the presence of oscillations alone is not a sufficient diagnostic. The confinement-related information lies in the {\it systematic dependence} of the oscillatory pattern on the longitudinal field $h_z$, at fixed system size and fixed quench protocol, and in its comparison with the non-confining case $h_z=0$. A clean lattice test \cite{Jiang:2025wpj}, verifies that the relevant oscillation frequency tracks the meson/confinement scale set by $h_z$, that the amplitude changes systematically with $h_z$. These features persist for times shorter than the finite-size recurrence time and remain stable as the system size is increased. Our holographic result should be compared with this confinement-controlled part of the spin-chain dynamics, not with generic finite-size recurrences.

The ultimate reason why oscillations occur (in holography) is that the radial motion of the
probe is bounded between a UV cutoff and a smooth end-of-space in the infrared, leading to
a periodic motion whose frequency is controlled by the scale of confinement. In the Ising model it is the finite Hilbert space (and spectrum) that is covered by the spread of the state.
\\
Let us be more precise about the analogy (rather than the comparison) we are trying to draw between these two different systems. The analogy with the Ising chain is based on the published analysis of Ref.~\cite{Jiang:2025wpj,Kormos_2016}. We have not performed an independent lattice calculation here. The Krylov quantity used in that comparison is state, or spread, complexity after a quench. In the ferromagnetic regime $h_x<1$, the theory at $h_z=0$ contains freely propagating domain walls, while turning on $h_z$ generates a linear potential between them and confines them into mesonic bound states. The quench protocol changes $h_z$, with the initial state taken to be in the pre-quenched theory. The analogy with our holographic setup is therefore {\it qualitative}: the holographic initial condition describes a localized heavy excitation released from the UV region of a confining geometry, while the spin-chain quench creates a state with overlap on the confined mesonic sector. We do not claim a one-to-one matching of initial states or a quantitative equality of the corresponding complexities. 
\\
As stated, these coincident behaviours are {\it qualitative}. It is beyond reasonable to search for quantitative coincidences, as both physical systems are very different. But, we find compelling the similarity of behaviour. Needless to say, our system has degrees of freedom that the Ising model does not display, like the R-symmetry conserved number of Section \ref{sec:2.3}, but one may imagine the existence of other condensed matter systems with qualitatively similar freedom.

Probably this should be further studied and we leave this for future investigations. Now, we study the Krylov complexity is various other holographic models dual to confining field theories in $(3+1)$-dimensions.

\section{Other confining models}\label{section3-othermodels}

In this section we study the complexity of different confining field theories, using their holographic duals. 
These models constitute canonical examples of confining gauge-string duals. They illuminate universal mechanisms underlying confinement, mass gap generation, and symmetry breaking, while also clarifying the geometric origin of strongly coupled infrared dynamics. As such, they continue to serve as benchmarks for both conceptual developments in holography and the systematic exploration of strongly coupled gauge theories.

Below we perform a systematic study of radial geodesics, proper momentum and complexity in a variety of Type II backgrounds dual to confining field theories. Even when the section seems a bit repetitive, we prioritised the presentation of some details.

To beging we develop generic expressions for the geodesic, proper momentum and complexity, for a generic background. We then apply these formulas to various holographic set-ups for confining theories. The models we study are: Witten’s D4-brane construction compactified on a circle with supersymmetry-breaking boundary conditions provides one of the earliest holographic realizations of four-dimensional Yang–Mills–like dynamics, exhibiting a mass gap, confinement, and linear quark–antiquark potentials \cite{Witten:1998zw}, see also \cite{Aharony:2006da} for a more recent summary. The addition of fundamental fields (quarks) was famously considered for this model in \cite{Sakai:2004cn}, see also \cite{Bigazzi:2014qsa}. The Klebanov–Strassler (KS) solution \cite{Klebanov:1998hh,Klebanov:2000nc, Klebanov:2000hb, Gubser:2004qj,Dymarsky:2005xt}, based on fractional D3-branes on the deformed conifold, offers a smooth, fully back-reacted geometry that captures confinement through a cascade of dualities and chiral symmetry breaking, closely mirroring some qualitative features of QCD-like theories. A version of the Klebanov-Strassler model that explores the moduli space of vacua along a purely baryonic branch of it was constructed by Butti, Gra\~na, Minasian, Petrini and Zaffaroni in \cite{Butti:2004pk}. The addition of flavours to KS was considered in \cite{Benini:2006hh, Benini:2007gx, Bigazzi:2008ie, Bigazzi:2008qq, Bigazzi:2008zt, Benini:2011cm}.  The background arising from D5-branes wrapped on an 
two-sphere inside the resolved conifold \cite{Maldacena:2000yy, Chamseddine:1997nm},
 provides a complementary realization of confinement tied to geometric twisting, with an explicit holographic interpretation of gaugino condensation \cite{Apreda:2001qb, DiVecchia:2002ks}. Fundamental fields were considered in this model, see \cite{Casero:2006pt, Casero:2007jj, Hoyos-Badajoz:2008znk, Nunez:2010sf}. An interesting geometric and field theoretical connection between models of wrapped D5 branes and the Baryonic branch of KS was developed in \cite{Maldacena:2009mw, Gaillard:2010qg, Caceres:2011zn}.

In the following we study the complexity of these QFTs  using their holographic duals. First, we develop generic expressions for the massive geodesic, equation of motion, its solution by quadratures (using conservation laws), and the holographic Krylov complexity. These expressions are valid for the simple geodesic studied here. More elaborated situations (like massive particles with global symmetry charges or geodesics exploring the internal space) should be considered separately. After deriving the generic expressions in Section \ref{sec:3.1}, we discuss the complexity in the various models summarised above, presenting a numerical analysis of the relevant equations.

\subsection{Generic expressions}\label{sec:3.1}
We  present general expressions that become useful in the coming sections, when evaluating the complexity for particular models.
We consider a massive geodesic described by the coordinate $r$ parametrised in terms of the time coordinate, $r(t)$. We do not excite any other coordinates (denoted by $\vec{v}$ below). The Einstein frame metric reads,
\begin{equation}
 ds_E^2= -A(r) dt^2+ A(r) B(r) dr^2+g_{ij}(r,\vec{v}) dv^i dv^j.\label{dsE}   
\end{equation}
As explained the $\vec{v}$-coordinates are not relevant for the geodesic here studied.
The action of a particle of mass $m$ that moves along the radial direction is,
\begin{equation}
 S=-m \int dt \sqrt{A(r)\left( 1-B(r)~\dot{r}^2\right)}.
 %, ~~A(r)= e^{\frac{\Phi}{2}}\hat{h}^{-\frac{1}{2}}, ~~B(r)= \hat{h} e^{2k}
 \label{action-KS}   
\end{equation}
The equation of motion is 
\begin{equation}
 \frac{d}{dt}\Bigg[ \frac{A B \dot{r}}{L}\Bigg]= \frac{-A'(1- B\dot{r}^2) -A B'\dot{r}^2}{2L},~~L= -m \sqrt{A(r)\left( 1-B(r)~\dot{r}^2\right)}.\label{eq-mot-r}  
\end{equation}
We denoted $A'=\frac{d A(r)}{dr}, B'=\frac{dB(r)}{dr}$. The momentum and the conserved Hamiltonian are,
\begin{equation}
P_r= \frac{m A B \dot{r}}{\sqrt{A(1- B \dot{r}^2)}},~~~  H=m \frac{A}{\sqrt{A(1- B \dot{r}^2)}}. \label{lionel} 
\end{equation}
From the conserved Hamiltonian we find,
\begin{eqnarray}
& &  \dot{r}=\pm\frac{1}{H}\sqrt{\frac{H^2-m^2A(r)}{B(r)}},\label{eq-r} \\
& & \pm \frac{(t-t_0)}{H}=\int_{r_{UV}}^{r} dr~\sqrt{\frac{B(r)}{H^2- m^2 A(r)}}.\label{t-de-r}
\end{eqnarray}
One can check that the first order equation (\ref{eq-r}) solves the second order equation of motion (\ref{eq-mot-r}). 

The complexity needs us to define the proper-coordinate $\bar{y}$ such that setting $\Delta t=\Delta v_i=0$ the metric reads
\begin{eqnarray}
& &ds_E^2= A(r) B(r)dr^2= d{\bar{y}}^2 \Longrightarrow d\bar{y}=\pm dr~\sqrt{A(r)B(r)}.\label{changetoybar}   \\
& & \to  \frac{d\dot{r}}{d\dot{\bar{y}}}= \frac{1}{\sqrt{AB}}.\nonumber
\end{eqnarray}
Following \cite{Caputa:2024sux} and {\it assuming} that the identification between proper momentum and  the time derivative of the complexity $\dot{C}(t)$ is valid, we have
\begin{eqnarray}
& &-\epsilon \dot{C}(t)= P_{\bar{y}}=\frac{\partial L}{\partial \dot{r}} \frac{d\dot{r}}{d\dot{\bar{y}}}= 
\frac{m A B  ~\dot{r}}{\sqrt{A (1- B \dot{r}^2)} }\times\frac{1}{\sqrt{A B}}= m \dot{r} \sqrt{\frac{B}{1- B\dot{r}^2}}~.\label{complexityBBgeneric}
\end{eqnarray}
We can use the Hamiltonian in eq.(\ref{lionel})
and the fact that at $t=0$ the position of the particle is $r_{UV}$ (being the velocity $\dot{r}(t=0)=0$) to set $H= m\sqrt{A(r_{UV})}$. From this find,
\begin{equation}
 \dot{r}=\pm \sqrt{\frac{1}{B(r)}\left[1-\frac{A(r)}{A(r_{UV})} \right]}, ~~~1- B(r) \dot{r}^2=   \frac{A(r)}{A(r_{UV})}.
\end{equation}
Putting these together, we find
\begin{equation}
 \dot{C}(t)=- \frac{m}{\epsilon}\sqrt{\frac{A(r_{UV})}{A(r)} -1}.   \label{diegoa}
\end{equation}
For this simpler expression to be useful, $r(t)$ must be found first. We emphasise that the above derivation assumes that at $t=0$, $r(0)=r_{UV}$ and the initial velocity vanishes. 

We can expand $r(t)$ close to boundary, $r(0)\equiv r_{UV}$, with $r_{UV}$ being the solution to $m A^{1/2}(r_{UV})=H$. The result is 
\begin{align} \label{eq:rUV}
    r(t)= r_{UV} - \frac{A'(r_{UV})}{4 A(r_{UV})B(r_{UV})}t^2 + O(t)^3,
\end{align}
The expansions of the complexity close to the boundary %and end of space 
reads
\begin{align} \label{eq:pyuv}
 \partial_t C(t)   \propto &P_{\bar y}\Bigg|_{t\sim0} =- \frac{A'(r_{UV})}{2 \sqrt{B(r_{UV})}A(r_{UV})} t +O(t)^{2}\;.
 % &  P_{\bar\rho}\Bigg|_{t\sim t_e} = \pm\frac{\sqrt{H^2-(Q/l)}}{\sqrt{Q/l}} +O(t-t_e)^{2}\;. 
\end{align}

In the forthcoming sections, we study the expressions for $r(t)$ in eq.(\ref{t-de-r}) and the complexity in (\ref{complexityBBgeneric}) for the various confining models discussed above. We study geodesics in Einstein frame, as we are discussing the motion of particles.

\subsubsection{Witten's model for Yang-Mills: D4 branes on $S^1$}
\begin{figure}
    \centering
    \includegraphics[width=0.5\linewidth]{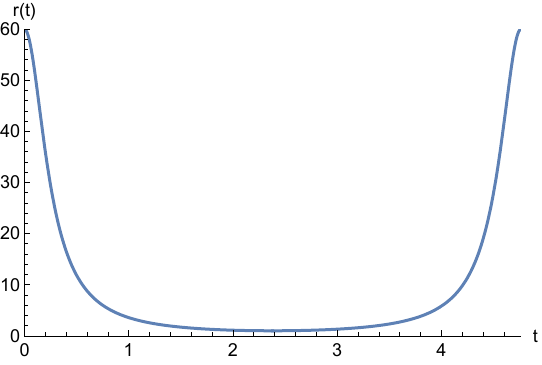}
        \includegraphics[width=0.53\linewidth]{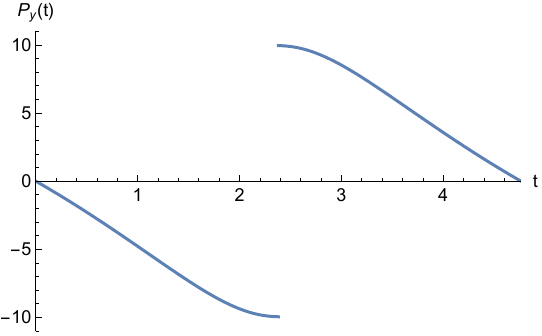}
             \includegraphics[width=0.5\linewidth]{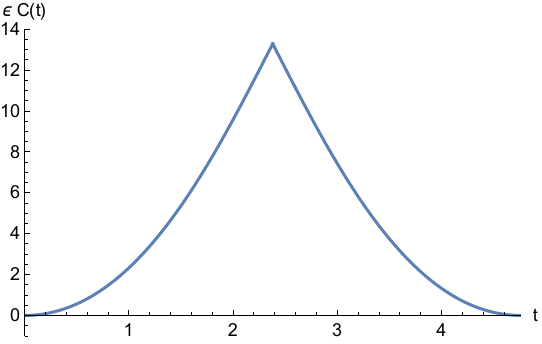}
    \caption{$r(t)$, $P_{y}$ and $C(t)$  for D4 branes on $S^1$ with $r_\Lambda=1,  N_c=10$ and $H=10$, $m=1$.}
    \label{fig:rtD4}
\end{figure}

The Einstein frame metric and dilaton corresponding to this background are,
\begin{eqnarray}
 & & ds_{E}^2= e^{-\frac{\Phi}{2} }\left( \frac{r}{L}\right)^{\frac{3}{2}}\Bigg[ -dt^2+d\vec{x}_3^2 +f(r) d\varphi^2 + \left( \frac{L}{r}\right)^3\left(\frac{dr^2}{f(r)}+r^2d\Omega_4\right)\Bigg],\label{metricD4onS1}\\
 & & e^{4\Phi}=g_s \left(\frac{r}{L}\right)^3, ~~f(r)=1-\left( \frac{r_\Lambda}{r}\right)^3,~~L^3=\pi g_s N_c\alpha'^{\frac{3}{2}},~~F_4\sim \text{Vol}_{S^4}.\nonumber
\end{eqnarray}
There is also a Ramond four-form proportional to the volume of the four-sphere that we do not quote in detail here.
Comparing with eq.(\ref{dsE}) and the other expressions in Section \ref{sec:3.1} we find,
\begin{eqnarray}
   & & A= e^{-\frac{\Phi}{2}}\left( \frac{r}{L}\right)^{\frac{3}{2}}, ~~B= \frac{L^3}{r^3 {f(r)}},\label{ABD4S1}\\
   & &\pm H L^{\frac{3}{2}}~\dot{r}=\sqrt{r^3{f(r)}\left[H^2 -m^2e^{-\frac{\Phi}{2}}\left( \frac{r}{L}\right)^{\frac{3}{2}} \right]},~~\dot{C}=-\frac{m}{\epsilon} \dot{r}\sqrt{\frac{L^3}{r^3-L^3 {f(r)}~\dot{r}^2}}.\label{quantitiesD4s1}
\end{eqnarray}
As we are unable to analytically solve the equation for $\dot{r}(t)$, we plot the numerical result for the position $r(t)$ and the complexity in Figure \ref{fig:rtD4}.

\subsubsection{Klebanov-Witten (KW) and Klebanov-Tseytlin (KT) models}\label{sectionKT}
In this section we study the holographic complexity using two different type IIB solutions: the Klebanov-Witten (KW) \cite{Klebanov:1998hh} and the Klebanov-Tseytlin (KT)  \cite{Klebanov:2000nc} backgrounds. Even when these are not dual to confining field theories, they serve as a path towards the Klebanov Strassler background, studied later. The KW background is dual to a four dimensional ${\cal N}=1$ SCFT consisting of two $SU(N)$ gauge nodes joined by bifundamental chiral superfields $A_1, A_2, B_1, B_2$ and a superpotential $W=k\epsilon_{ij}\epsilon_{ab}A_iB_aA_jB_b$. The SCFT is strongly coupled (the anomalous dimension of the chiral multiplets is $\gamma=-\frac{1}{2}$). The KT background can be thought as the holographic dual to the quasi-marginal deformation of the KW SCFT. An inbalance between the gauge nodes is introduced, leading to a quiver of the form $SU(N)\times SU(N+M)$. The parameter $M$ is represented holographically by the addition of fractional D3 branes that trigger an RG-flow. With approximately the same anomalous dimensions to the KW ones, one finds the beta functions and anomalies of each gauge group (all proportional to $M$). The QFT dynamics is beautifully explained in \cite{Strassler:2005qs}.

Both backgrounds have an Einstein frame metric (the dilaton is vanishing in both cases) that reads
\begin{eqnarray}
 & &ds_E^2=\hat{h}^{-\frac{1}{2}}\left(-dt^2+d\vec{x}_3^2 \right) + \hat{h}^{\frac{1}{2}}\left( dr^2+ r^2 ds^2_{T^{1,1}}\right),\label{metricaKWKT}\\
 & &ds^2_{T^{1,1}}= \frac{1}{6} (d\theta_1^2+\sin^2\theta_1 d\varphi_1^2)+ \frac{1}{6} (d\theta_2^2+\sin^2\theta_2 d\varphi_2^2)+\frac{1}{9}\left(d\psi+\cos\theta_1 d\varphi_1+\cos\theta_2 d\varphi_2 \right)^2.\nonumber\\
 & & \text{For KW}: \hat{h}_{KW}=\frac{Q^4}{r^4},~\text{for KT}: \hat{h}_{KT}=\frac{Q^4}{r^4}\times\! \frac{3 (g_s M)^2}{2\pi}\!\log(\frac{r}{r_0}).\nonumber
\end{eqnarray}
The KW background is completed with a RR five form. The KT background requires a five form, a Ramond three form and a Neveu-Schwarz three form. These forms are not quoted here as they are not needed in our calculation.

Comparing with the expression in Section \ref{sec:3.1}
we find,
\begin{eqnarray}
    & & A=\hat{h}^{-\frac{1}{2}},~~~B=\hat{h}.
\end{eqnarray}
Where $\hat{h}$ stands for the warp factors $\hat{h}_{KW}$ or $\hat{h}_{KT}$ respectively.

For each background we can use the generic expressions
in Section \ref{sec:3.1}. For the KW background we find,
\begin{eqnarray}
%& &S= -m \int dt\sqrt{\frac{r^2}{Q^2}\left(1-\frac{Q^4}{r^4}\dot{r}^2\right)},\nonumber\\
& & \dot{r}=\pm\frac{r^2}{HQ^3}\sqrt{H^2 Q^2- m^2 r^2}\to r(t)=\frac{H Q^2}{\sqrt{m^2 Q^2 + H^2(t-t_0)^2}},\nonumber\\
& & \dot{C}(t)= -\frac{P_{\bar{y}}}{\epsilon}\sim- \frac{H}{Q}(t-t_0).\label{resultT11}
\end{eqnarray}
This reproduces the conformal result derived in  \cite{Caputa:2024sux} for a generic 2d CFT. We see that this result is universal for CFTs in generic dimension.

\begin{figure}
    \centering
    \includegraphics[width=0.5\linewidth]{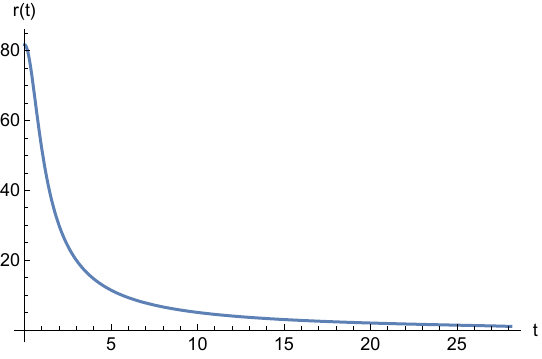}
        \includegraphics[width=0.53\linewidth]{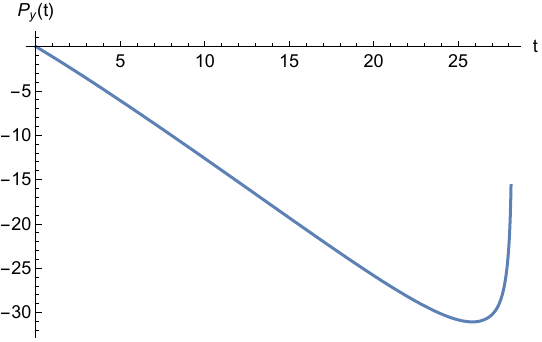}
             \includegraphics[width=0.5\linewidth]{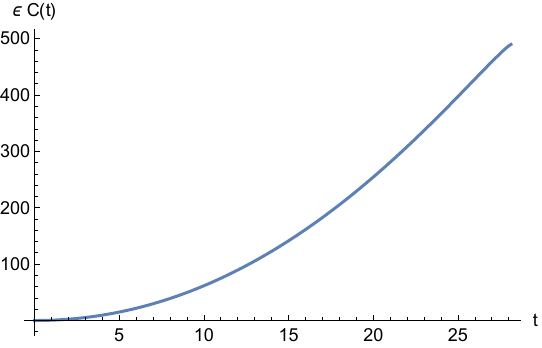}
    \caption{$r(t)$, $P_{y}$ and $C(t)$  for KT solution with $r_0=1,  M=10$ and $H=10$, and $m=1$.}
    \label{fig:rtKT}
\end{figure}

In the case of KT background
the quantities relevant for the complexity calculation are,
\begin{eqnarray}
 & & \pm\frac{(t-t_0)}{H}= \int_r^\infty \sqrt{\frac{\hat{h}_{KT}^{\frac{3}{2}}}{H^2~\hat{h}_{KT}^{\frac{1}{2} }- m^2}},\label{rdotKT}\\
 & &  \dot{C}(t)=-\frac{P_{\bar{y}}}{\epsilon}\sim - m \dot{r}\sqrt{\frac{\hat{h}_{KT}}{1-\hat{h}_{KT}~\dot{r}^2}}. 
 \label{complexityKT}
\end{eqnarray}
The problem seems analytically intractable. We present a numerical evaluation of $r(t)$ from eq.(\ref{rdotKT}), then use it to find the complexity in Figure \ref{fig:rtKT}. 
% \begin{figure}
%     \centering
%     \includegraphics[width=0.5\linewidth]{rtKT.pdf}
%         \includegraphics[width=0.5\linewidth]{CtKT.pdf}
%     \caption{$r(t)$ and $P_{y}$ for KT with $r_0=1,  M=10$}
%     \label{fig:rtKT}
% \end{figure}

It must be emphasised that for the KT background the result of the geodesic and the associated complexity are {\it not} trustable close to the end of the space $r=r_0$--see the expression for $\hat{h}_{KT}$ in eq.(\ref{metricaKWKT}). In fact, the KT background is singular at that point. The KT background does {\it not} capture the Physics of confinement and does not strictly belong to the set of models we study here. The results in KT should be trustable far away from the singularity. Only  close to the UV-region of the flow $r=r_{UV}$ we can trust the result. 
The monotonic behaviour shown for the KT background should not be interpreted as evidence that oscillations are impossible in the corresponding cascading theory.  The geodesic calculation in KT is trustworthy only in the region where the KT approximation is valid, namely sufficiently far from the singularity and close to the UV part of the flow. Since the smooth IR end of space is absent from the KT geometry, the mechanism producing bounded radial motion and oscillatory complexity is also absent. Nevertheless, one can think about singular geometries, for which an oscillatory behaviour has been observed, see \cite{Zoakos:2026obl}.

In what follows we discuss the Klebanov-Strassler background and its Baryonic Branch, that amend the previous singularity, precisely incorporating the phenomena of confinement and symmetry breaking.

\subsubsection{Klebanov-Strassler (KS) model}\label{sectionKS}
The Klebanov-Strassler field theory is (roughly) the 'IR-completion' of the Klebanov-Tseytlin QFT. Whilst KT is dual to a sequence of Seiberg dualities, this description ceases to be valid at some low energy scale, see \cite{Strassler:2005qs, Dymarsky:2005xt}. The KS field theory takes over, incorporating confinement and symmetry breaking in the dynamics. In more detail, KS describes the evolution of the two-nodes quiver $SU(N)\times SU(N+M)$ in the situation $N=k M$ (being $k$ a natural number, typically large).
In this fine-tuned case the last step of the cascade leads to a version of SQCD with number of flavours equal to the number of colours. The low energy dynamics is described in \cite{Seiberg:1994bz, Dymarsky:2005xt}, leading to confinement and the spontaneous breaking of baryonic symmetry (there is the possibility of mesonic branches, not captured by the KS background). The KS QFT is tuned to the case in which the VEV of the baryon and anti-baryon operators are equal (a restriction lifted in the next section, when we discuss the baryonic branch of KS).

The backgrounds we discuss in this section and in Sections  \ref{sectionBB},  \ref{sectionD5S2}, can be written as particular cases of the generic background that in Einstein frame reads,
\begin{eqnarray}
& & ds^2_{E}=e^{\frac{\Phi}{2}}\hat{h}^{-\frac{1}{2}}\left[ -dt^2+ dx_1^2+dx_2^2+dx_3^2\right] + e^{\frac{\Phi}{2}}\hat{h}^{\frac{1}{2}}\big[e^{2k}dr^2+ e^{2h}(d\theta^2+\sin^2\theta d\varphi^2)+\nonumber\\
& & \frac{e^{2g}}{4}\left[(\omega_1+a d\theta)^2+(\omega_2-a\sin\theta d\varphi)^2 \right]+
\frac{e^{2k}}{4}(\omega_3-\cos\theta d\varphi)^2\big],  ~~~\Phi=\Phi(r). \label{backgroundBB} 
\end{eqnarray}
We have used the left invariant forms of $SU(2)$,
\begin{equation}
\omega_1=\cos\psi_2 d\theta_2+ \sin\psi_2\sin\theta_2 d\varphi_2, ~~\omega_2= -\sin\psi_2 d\theta_2+ \cos\psi_2\sin\theta_2 d\varphi_2,~~\omega_3=d\psi_2 +\cos\theta_2d\varphi_2.\nonumber   
\end{equation}
The functions $[\Phi,k,h,g,a,\hat{h}]$ depend only on the radial coordinate. Their expressions are written in terms of two functions $P(r)$ and $Q(r)$ \cite{Hoyos-Badajoz:2008znk, Nunez:2008wi}. For a summary of the notation and derivation, see Appendix A of \cite{Conde:2011aa}.
\begin{eqnarray}
& & Q=N_c(2r\,\coth(2r)-1)\,, ~~e^{2h}\,=\,\frac{1}{4}\,
\frac{P^2-Q^2}{P\coth(2r)-Q}\,,~~~ e^{2g}\,=\,P\,\coth(2r)\,-\,Q\,,\label{xxxxdd1.A}\\
& & e^{2k}=\frac{P'}{2}\,,~~~ a=\frac{P}{P\cosh(2r)-Q\sinh(2r)}\,,~~~ e^{4\Phi-4\Phi_0}= \frac{2\sinh(2r)^2}{(P^2-Q^2)P'}\,, ~~\hat{h}= 1-\kappa^2 e^{2\Phi}.\nonumber
\end{eqnarray}
The function
$P(r)$ is given as the solution of the following ``master equation''
\begin{equation}
P'' + P'\Big(\frac{P'+Q'}{P-Q} +\frac{P'-Q'}{P+Q} - 4 
\coth(2r)\Big)=0\,.
\label{master.A}
\end{equation}
Different solutions of the master equation (\ref{master.A}) were studied in \cite{Nunez:2008wi,Gaillard:2010qg}. The background is complemented by RR three and five form field strengths and a NS three form. As above, these are not needed for our calculation and not quoted here.

The case of the KS background the relevant functions are
\begin{eqnarray}
& & e^{\Phi}=1, ~~e^{2k}=\frac{\epsilon^{\frac{4}{3}}}{6 {\cal K}(r)^2}, ~~ A= \hat{h}^{-\frac{1}{2}}, ~~~B=\frac{\hat{h}~\epsilon^{\frac{4}{3}}}{6~ {\cal K}^2(r)},\nonumber\\
& & {\cal K}^2(r) =\frac{\left( \sinh(2r) -2r\right)^{\frac{2}{3}}}{2^{\frac{2}{3}} \sinh^2(r)},\nonumber\\
& & \hat{h}(r)= (2^{\frac{1}{3}} g_s M\alpha' \epsilon^{-\frac{4}{3}})^2 \int_r^\infty dx\left( \frac{x \coth x-1}{\sinh^2 x} \right) \left(\sinh(2x)- 2x\right)^{\frac{1}{3}}.\label{functionsKS}
\end{eqnarray}
For the expression for the function $P(r)$ that gives $\hat{h}_{KS}$, see Section 2.1.1 in \cite{Conde:2011aa}. We do not make use of this here as the expressions to calculate the complexity can be obtained using eq.(\ref{functionsKS}).

\begin{figure}
    \centering
    \includegraphics[width=0.5\linewidth]{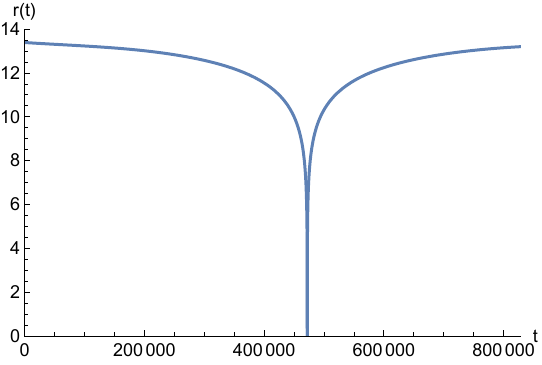}
        \includegraphics[width=0.53\linewidth]{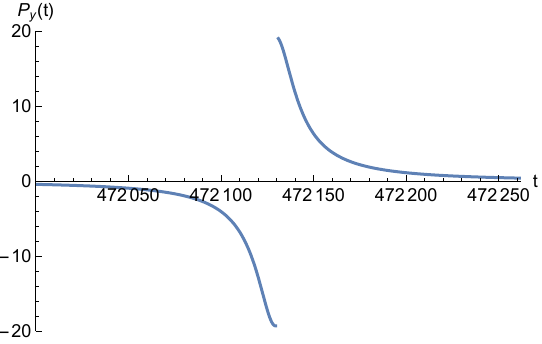}
             \includegraphics[width=0.5\linewidth]{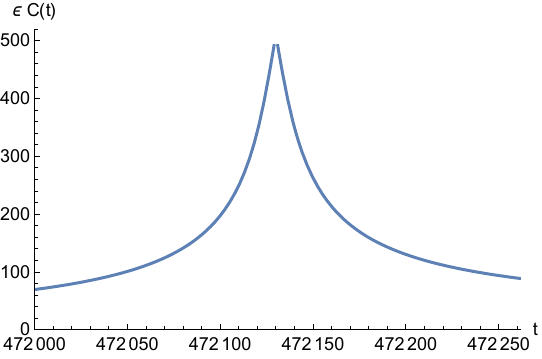}
    \caption{$r(t)$, $P_{y}$ and $C(t)$  for KS solution with $\epsilon=1, M=10$, $H=10$, and $m=1$.}
    \label{fig:rtKS}
\end{figure}

In fact, to calculate the complexity, we need to solve for $\dot{r}(t)$ for the geodesic. The results are,
\begin{eqnarray}
& &   \pm\frac{(t-t_0)}{H}= \int_r^\infty \sqrt{\frac{\hat{h}^{\frac{3}{2}} \epsilon^{\frac{3}{4}}}{6\left(H^2\hat{h}^{\frac{1}{2} }- m^2\right){\cal K}^2(r)}},\label{rdotKS}\\
 & & \dot{C}(t)=-\frac{P_{\bar{y}}}{\epsilon}=
  -\frac{m}{\epsilon}\dot{r}\sqrt{\frac{\hat{h}_{KS} \epsilon^{\frac{4}{3}}}{6 {\cal K}^2 -\hat{h}_{KS} \epsilon^{\frac{4}{3}} \dot{r}^2}}=-\frac{m}{\epsilon}\sqrt{\sqrt{\frac{\hat{h}_{KS}(r)}{\hat{h}_{KS}(r_{UV})}} -1}.
\end{eqnarray}
We used eq.(\ref{diegoa}) in the last equality. 
In Figure \ref{fig:rtKS} we find the function $r(t)$, the momentum and the proper momentum, together with the complexity for the KS background. 

\subsubsection{The Baryonic Branch of KS}\label{sectionBB}
In this section we study the supergravity background written by Butti, Gra\~na, Minasian, Petrini and Zaffaroni \cite{Butti:2004pk}. This background describes the possibility of exploring baryonic vacua in the very last step of the KS cascade. A nice description of the field theory is given in \cite{Gubser:2004tf}. The holographic dual background is of the form in eq.(\ref{backgroundBB}). In particular, the function $P(r)$ is not known analytically, but can be expressed in terms of its UV ($r\to\infty$) and IR ($r\to 0$) expansions, with a numerical interpolation between the two expansions, that is easily found \footnote{There is an analytic way of writing solutions in terms of integrals, discussed in \cite{Nunez:2008wi, Hoyos-Badajoz:2008znk,Gaillard:2010qg, Conde:2011aa}}. In more detail,
for $r\to\infty$ we find an expansion
\begin{eqnarray}
&P&=e^{4r/3}\Big[ c_+  
+\frac{e^{-8r/3} N_c^2}{c_+}\left(
4r^2 - 4r +\frac{13}{4} \right)+ e^{-4r}\left(
c_- -\frac{8c_+}{3}r \right)+\nonumber\\
& & + \frac{N_c^4 e^{-16r/3}}{c_+^3}
\left(\frac{18567}{512}+\frac{2781}{32}r +\frac{27}{4}r^2 +36r^3\right)  + {\cal O}(e^{-20r/3})
\Big]\,.
\label{UV-II-N}
\end{eqnarray}
Notice that this expansion 
involves two integration constants, $c_+>0$ and $c_-$. 
The background functions $[h,g,k,a,\Phi, \hat{h}]$
at large $r$ are written for reference
in Appendix A of \cite{Conde:2011aa}.
Regarding the IR expansion, 
we look for solutions with $P\to 0$ 
as $r\to 0$, 
in which case we find
\begin{equation}
P= h_1 r+ \frac{4 h_1}{15}\left(1-\frac{4 N_c^2}{h_1^2}\right)r^3
+\frac{16 h_1}{525}\left(1-\frac{4N_c^2}{3h_1^2}-
\frac{32N_c^4}{3h_1^4}\right)r^5+{\cal O}(r^7)\,,
\label{P-IR}\end{equation}
where $h_1>2N_c$ is again an arbitrary constant; 
there is another integration constant, $P(0)$,
taken to zero here, to avoid singularities \footnote{Solutions with large $P(0)$ are referred as  'walking' and have been used to study QFTs whose beta function 'running' is very slow, see \cite{Elander:2011mh, Nunez:2008wi}}. The IR expansion of the background functions $[h,g,k,a, \Phi,\hat{h}]$ are given in Appendix A of  \cite{Conde:2011aa}.

%To match the UV expansion amounts to taking as well $c_-=0$ in \eqref{UV-II-N}.
%This gives background functions
%that are quoted in Appendix A of \cite{Conde:2011aa}.
There is a smooth numerical interpolation between both 
expansions. There is  only one independent 
parameter; given a value for one of the constants $\{c_+, h_1\}$, 
the requirement that the solution matches \emph{both} 
expansions is sufficient to determine the value of the other integration constant.

\begin{figure}
    \centering
    \includegraphics[width=0.5\linewidth]{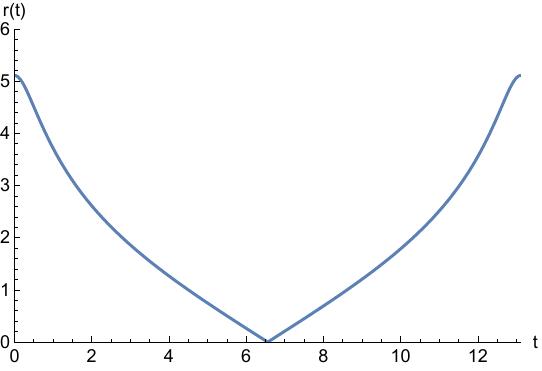}
        \includegraphics[width=0.53\linewidth]{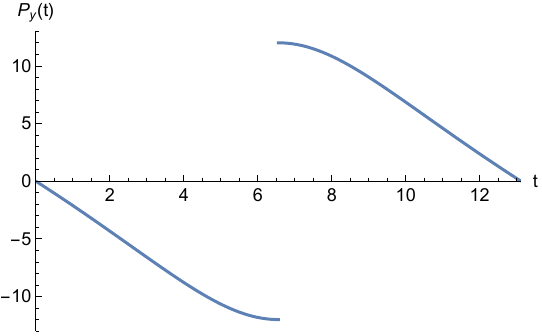}
             \includegraphics[width=0.5\linewidth]{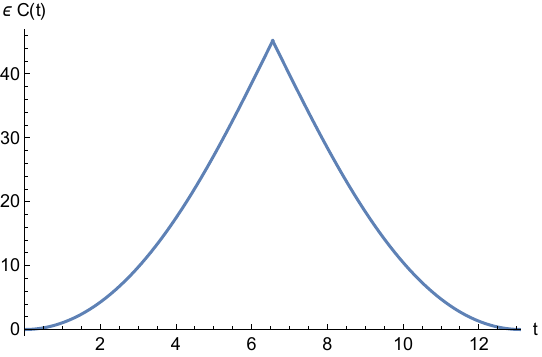}
    \caption{$r(t)$, $P_{y}$ and $C(t)$  for the baryonic branch of the KS with $h_1=4N_c, N_c=10$, $H=10$, $m=1$.}
    \label{fig:rtBa}
\end{figure}

In Figure \ref{fig:rtBa} we find the function $r(t)$, the momentum and the proper momentum, together with the complexity for this background. 

\subsubsection{D5 branes on  on $S^2$}\label{sectionD5S2}
The field theory corresponding to a stack of $N$ D5 branes that wrap a two cycle inside the resolved conifold is (at low energies compared with the size of the cycle) a four dimensional version of ${\cal N}=1$ SYM, complemented by a tower of massive chiral and vector multiplets. The weakly coupled Lagrangian is written in \cite{Andrews:2005cv, Andrews:2006aw, Hoyos-Badajoz:2008znk}. The holographic dual written in \cite{Chamseddine:1997nm, Maldacena:2000yy} is of the form in eq.(\ref{backgroundBB}), with $P(r)=2 N r$. This is an analytic solution of the master equation (\ref{master.A}). The background dual to the QFT above described is smooth and asymptotes (for large $r$) to the background of $N$ D5 branes, indicating that the 4d QFT gets UV completed by the tower of KK modes to the ${\cal N}=(1,1)$ LST.

\begin{figure}
    \centering
    \includegraphics[width=0.5\linewidth]{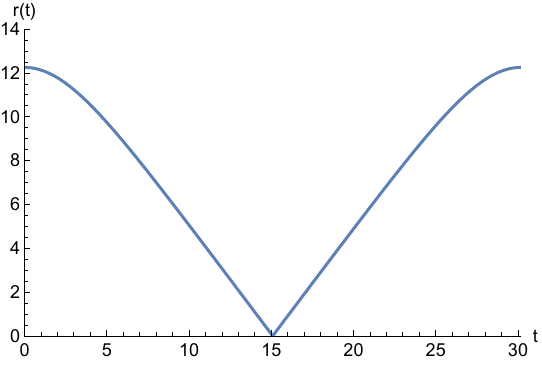}
        \includegraphics[width=0.53\linewidth]{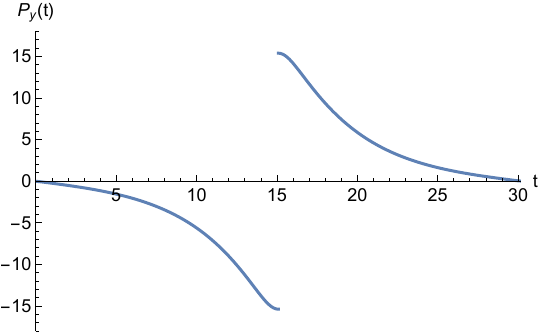}
             \includegraphics[width=0.5\linewidth]{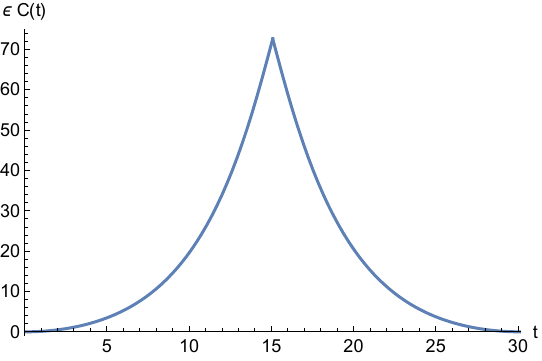}
    \caption{$r(t)$, $P_{y}$ and $C(t)$  for the D5 solution with $N_c=10$, $m=1$ and $H=10$.}
    \label{fig:rtD5}
\end{figure}

We calculate with $P=2 Nr $ and $\hat{h}=1$ (or $\kappa=0$). We have $A=e^{\frac{\Phi}{2}}$ and $B=1$.
The expressions for $\dot{r}$ and the derivative of the complexity are,
\begin{eqnarray}
 & &\dot{r}=\pm\frac{1}{H}\sqrt{H^2- m^2 e^{\frac{\Phi}{2}}}, ~~\frac{(t-t_0)}{H}=\int_{r_{UV}}^r dr~\sqrt{\frac{1}{H^2- m^2 e^{\frac{\Phi}{2}}}},\\
 & & \dot{C}= -\frac{m\dot{r}}{\epsilon\sqrt{1-\dot{r}^2}}= -\frac{m}{\epsilon}\sqrt{\frac{e^{\frac{\Phi(r_{UV})}{2}}}{e^{\frac{\Phi(r)}{2} }} -1}.
\end{eqnarray}
We have used eq.(\ref{diegoa}) in the last equality.
In Figure \ref{fig:rtD5} we find the function $r(t)$, the momentum and the proper momentum, together with the complexity for this background.  Appendix \ref{lastapp} studies the case of a SCFT coupled to gravity and its complexity.
\\
\underline{ \bf Summary of this section}: in all these confining models we find the oscillations in the complexity found in the Anabal\'on-Ross model. The oscillatory complexity seems to be a universal characteristic of confining systems. It would be nice to study the addition of conserved R-charges to the study in this section. The oscillatory behaviour should be viewed as the universal, coarse feature common to smooth holographic confining geometries. The detailed waveform of the oscillatory motion contains more refined information. For a radial trajectory in a metric of the form
\begin{equation*}
 ds_E^2=-A(r)dt^2+A(r)B(r)dr^2+\cdots,
\end{equation*}
with initial point $r_{\rm UV}$, the period and amplitude are controlled by integrals and ratios involving the warp factors $A(r)$ and $B(r)$. Therefore different geometries need not give identical complexity curves. The Anabalon--Ross model, Witten's D4 model, Klebanov--Strassler, the baryonic branch, and wrapped-D5 backgrounds differ in their running, their IR caps, their symmetry-breaking data, and their available charge sectors. These differences are reflected in the period, amplitude, anharmonicity and charge dependence of $C_K(t)$. In this sense Krylov spread complexity is not merely a binary detector of confinement. It can also act as a quantum-information probe of model-dependent IR structure. A systematic reconstruction of such IR data from the full complexity waveform is an interesting problem that we leave for future work.

\section{Conclusions and closing comments}\label{concl}
In this work we have performed a systematic holographic study of Krylov complexity in a wide class of confining quantum field theories. Using the geometric prescription relating the time derivative of the complexity to the proper momentum of a massive probe, we analysed radial geodesics in several top–down gravity duals that exhibit confinement and a mass gap. First, we treated in detail the Anabal\'on–Ross background, where analytic control is possible, and extended the analysis to configurations with non-trivial angular momentum and R-charge. We discuss qualitative and quantitative differences between these cases. We then developed general expressions for radial motion and proper momentum in arbitrary warped backgrounds and applied them to canonical confining models, including Witten’s D4 construction, Klebanov–Strassler, its baryonic branch, and wrapped D5–brane systems.

Across all genuinely confining geometries we observe a  robust qualitative feature: the Krylov complexity exhibits oscillatory behaviour. The origin of these oscillations is geometric and transparent in the holographic picture. The radial motion of the probe is bounded between a UV cutoff and a smooth end-of-space in the infrared, leading to a periodic motion whose frequency is controlled by the scale of confinement. The amplitude of the oscillations depends both on the UV cutoff and inversely on the confinement scale, while additional conserved charges modify the amplitude and frequency in a systematic manner. 

These features persist across very different supergravity constructions, strongly suggesting that they are not artefacts of a specific background but instead reflect universal properties of confining dynamics.
An illuminating aspect of our analysis is the qualitative comparison with the transverse-field Ising model in the presence of a longitudinal perturbation. In that system, turning on the longitudinal field produces confinement of domain walls and induces oscillations in Krylov complexity whose frequency increases and amplitude decreases with the confinement parameter. We find a compelling parallel: the holographic parameter controlling the end–of–space and hence the confinement scale plays a role analogous to the longitudinal field. In both systems, confinement reorganises the Hilbert space in a way that is sharply detected by Krylov complexity. Although quantitative matching between holographic gauge theories and the Ising model is neither expected nor attempted, the qualitative agreement strongly supports the idea that complexity is a sensitive probe of confinement beyond traditional diagnostics such as Wilson loops or spectral gaps.

Equally important is the contrast with non–confining geometries. In backgrounds that do not possess a smooth infrared cap, the oscillatory pattern is absent. This reinforces the interpretation that the bounded radial motion -- and therefore the presence of an intrinsic infrared scale -- is directly responsible for the observed behaviour of the complexity.

Our results open several promising directions for future research. A first natural step is to extract analytic expressions for the oscillation period in those cases where closed–form solutions are available,  making the dependence on the confinement scale completely explicit and comparable with Lattice results in other confining models. It would also be highly desirable to develop a field–theoretic understanding of these oscillations directly from the Lanczos coefficients in strongly coupled confining theories. Another important direction concerns subregion and operator–dependent notions of complexity, and whether similar oscillatory signatures persist in those frameworks. Finally, exploring the interplay between conserved global charges, symmetry breaking patterns, and complexity growth may reveal further universal structures in confining quantum systems.
Overall, our analysis provides strong evidence that Krylov complexity captures a universal and dynamically rich signature of confinement. The oscillatory behaviour  appears to be a geometric manifestation of the infrared structure of confining theories, offering a new bridge between holography, operator growth, and non–perturbative gauge dynamics.
\section*{Acknowledgments:} We wish to thank various colleagues for discussions and comments. We are grateful to: E. Caceres, D. Chatzis, X. Jiang, J. Halimeh ,W. M\"uck, H. Nastase, J. Pedraza, D. Roychowdhuri, J. Sim\'on,   N. Srivatsa, J. Subils, R. Terrazas, D. Zoakos

A. F. is supported by EPSRC's grant UKRI3028. C. N. is supported by STFC’s grants ST/Y509644-1, ST/X000648/1 and ST/T000813/1.

\appendix
\section{Exact results for $J\neq0$ case}\label{exactresulJ}
In this section, we consider the SUSY preserving situation $\mu=0$ with $J\neq0$.  The function $f(r)$ is given by $f(r)=1- \frac{Q^2l^2}{r^4}$. The space ends at $r_*=\sqrt{Ql}$. We launch the massive probe from $r_{UV}$, with zero initial velocity. 
To streamline the calculation, it is convenient to work in a  different coordinate $\frac{r}{l}= \frac{l}{z}$. In this coordinate, the space ranges between 
$[z_{UV}, z_*]=[z_0,\sqrt{\frac{l^3}{Q}}]$.

The integral in eq.\eqref{betterex} reads (in the $z$ coordinate and setting the integration constant $t_0=0$ and $l=1$),
\begin{align}
    &\frac{t}{H} =\int_{z_0}^z d\tilde z~ \frac{\tilde z}{{\sqrt{-H^2 Q^2 \tilde z^6+Q^2 \tilde z^4+(H^2-J^2) \tilde z^2-1}}}.\label{eq:Jz}
\end{align}
In a process similar to the one mentioned in section \ref{sec:ARQ}, one can solve the integral and obtain $t(z)$ and invert it to find $z(t)$ or $r(t)$ as
\begin{align}
     z(t)&= \sqrt{\frac{1}{Q}}\sqrt{\frac{b (-a + c) + (a - b) c\operatorname{sn}^2(\sqrt{a-c}  \sqrt{Q}\; t,k)}{-a + c + (a - b)\operatorname{sn}^2(\sqrt{a-c}  \sqrt{Q}\; t,k)}},\\   
     r(t)&=\sqrt{Q}\sqrt{\frac{-a + c + (a - b)\operatorname{sn}^2(\sqrt{a-c}  \sqrt{Q}\; t,k)}{b (-a + c) + (a - b) c\operatorname{sn}^2(\sqrt{a-c}  \sqrt{Q}\; t,k)}}.\label{ztJfinal}
\end{align}
Here the set of parameters is defined as
\begin{eqnarray}
    k^2=\frac{a-b}{a-c}, \quad g=\frac{2}{\sqrt{a-c}},
\end{eqnarray}
with
\begin{align}
a&=\frac{Q}{3 H^2}+\nonumber\\&\frac{\left(1-i \sqrt{3}\right) \sqrt[3]{18 H^4 Q+9 H^2 J^2 Q+\sqrt{\left(18 H^4 Q+9 H^2 J^2 Q-2 Q^3\right)^2+4 \left(-3 H^4+3 H^2 J^2-Q^2\right)^3}-2 Q^3}}{6 \sqrt[3]{2} H^2}-\nonumber\\&\frac{\left(1+i \sqrt{3}\right) \left(-3 H^4+3 H^2 J^2-Q^2\right)}{3\ 2^{2/3} H^2 \sqrt[3]{18 H^4 Q+9 H^2 J^2 Q+\sqrt{\left(18 H^4 Q+9 H^2 J^2 Q-2 Q^3\right)^2+4 \left(-3 H^4+3 H^2 J^2-Q^2\right)^3}-2 Q^3}},
\end{align}
\begin{align}
b&=\frac{Q}{3 H^2}+\nonumber\\&\frac{\left(1+i \sqrt{3}\right) \sqrt[3]{18 H^4 Q+9 H^2 J^2 Q+\sqrt{\left(18 H^4 Q+9 H^2 J^2 Q-2 Q^3\right)^2+4 \left(-3 H^4+3 H^2 J^2-Q^2\right)^3}-2 Q^3}}{6 \sqrt[3]{2} H^2}-\nonumber\\&\frac{\left(1-i \sqrt{3}\right) \left(-3 H^4+3 H^2 J^2-Q^2\right)}{3\ 2^{2/3} H^2 \sqrt[3]{18 H^4 Q+9 H^2 J^2 Q+\sqrt{\left(18 H^4 Q+9 H^2 J^2 Q-2 Q^3\right)^2+4 \left(-3 H^4+3 H^2 J^2-Q^2\right)^3}-2 Q^3}},
\end{align}
\begin{align}
    c&=\frac{Q}{3 H^2}-\nonumber\\&\frac{\sqrt[3]{18 H^4 Q+9 H^2 J^2 Q+\sqrt{\left(18 H^4 Q+9 H^2 J^2 Q-2 Q^3\right)^2+4 \left(-3 H^4+3 H^2 J^2-Q^2\right)^3}-2 Q^3}}{3 \sqrt[3]{2} H^2}+\nonumber\\&\frac{\sqrt[3]{2} \left(-3 H^4+3 H^2 J^2-Q^2\right)}{3 H^2 \sqrt[3]{18 H^4 Q+9 H^2 J^2 Q+\sqrt{\left(18 H^4 Q+9 H^2 J^2 Q-2 Q^3\right)^2+4 \left(-3 H^4+3 H^2 J^2-Q^2\right)^3}-2 Q^3}}.
\end{align}
We  expand $z(t)$ close to boundary (at position $z_{UV}\equiv z_0$), $z(t=0)=z_0= \frac{\sqrt{b}}{\sqrt{Q}}$ 
\begin{align} \label{eq:z0J}
    z(t)= \frac{\sqrt{b}}{\sqrt{Q}} +\frac{\sqrt{Q} (a-b) (b-c)}{2 \sqrt{b}}t^2+ \cdots,
\end{align}
and at the return point $z_*=z(t_e)= \sqrt{\frac{a}{Q}}$,%=\sqrt{\frac{l^{3}}{Q}}$.
\begin{eqnarray}  \label{eq:zteJ}
    & & z(t)= \sqrt{\frac{a}{Q}}-\frac{(a-b) (a-c)}{2 \sqrt{\frac{a}{Q}}}(t-t_e)^2 + \cdots.
\end{eqnarray}
It is interesting to compute the time to reach $r=r_*= \sqrt{\frac{Q}{a}}$. The result is
\begin{equation}
 t_e=\frac{K\left(\frac{a-b}{a-c}\right)}{\sqrt{Q} \sqrt{a-c}},\label{halfperiodJ}  
\end{equation}
where $K$ represents the Jacobi elliptic function. One can check that the frequency of the motion is inversely proportional to the parameter $J$ and the range of the motion $\sqrt{\frac{Q}{b}}-\sqrt{\frac{Q}{a}}$ is also inversely proportional to this parameter. 

The proper momentum $P_y$ close to the boundary and end of space are
\begin{align} \label{eq:pbarrhoexpa}
 \partial_t \mathcal{C}(t)   \propto &P_{\bar y}\Bigg|_{t\sim0} =-\frac{\sqrt{b H^2-Q}}{\sqrt{Q}}- \frac{H^2 \sqrt{Q} (a-b) (b-c)}{2 \sqrt{b H^2-Q}}t^2+O(t)^3\;. \\
 &  P_{\bar y}\Bigg|_{t\sim t_e} = \pm\frac{\sqrt{a H^2-Q}}{\sqrt{Q}}\mp\frac{\left(H^2 \sqrt{Q} (a-b) (a-c)\right) (t-t_e)}{2 \sqrt{a H^2-Q}}+O(t-t_e)^{3}\;. \label{complexitylargeta}
\end{align}
Note that the momentum at early times is not linear and is starting from a constant term. This is related to the fact that there is an initial momentum given to the particle. It is in the direction of $\phi$ and proportional to the parameter $J$.
\\
\underline{\it How symmetry resolution can plausibly give a short-time linear term?}
\\
Let $H=H^\dagger$ be an Hermitian hamiltonian and let $Q$ be a conserved charge,  that is$[H,Q]=0$. The operator $\Pi_q$ projects onto the Hilbert space of fixed charge $q$, denoted by $\mathcal H_q$.  For a generic initial state
\begin{equation}
 |\psi\rangle=\sum_q \sqrt{p_q}\,|\psi_q\rangle,\qquad
 |\psi_q\rangle=\frac{\Pi_q|\psi\rangle}{\sqrt{p_q}},\qquad
 p_q=\langle\psi|\Pi_q|\psi\rangle,
\end{equation}
there is a standard, sector-adapted Krylov problem. In fact, following \cite{Caputa:2025ozd, Caputa:2025mii}, we build $|K_n^{(q)}\rangle$ from
$|\psi_q\rangle,H_q|\psi_q\rangle,H_q^2|\psi_q\rangle,\ldots$, where
$H_q=\Pi_qH\Pi_q$ is Hermitian on $\mathcal H_q$.  Then
\begin{equation}
 |\psi_q(t)\rangle=e^{-iH_qt}|\psi_q\rangle
 =\sum_n \phi_n^{(q)}(t)|K_n^{(q)}\rangle,\qquad
 C_q(t)=\sum_n n\,|\phi_n^{(q)}(t)|^2 .
\end{equation}
Since $\phi_n^{(q)}(0)=\delta_{n0}$, the usual short-time result is {\it unavoidable},
\begin{equation}
 C_q(t)=\big(b_1^{(q)}\big)^2t^2+O(t^4),\qquad
 b_1^{(q)}=\sqrt{\langle H_q^2\rangle_q-\langle H_q\rangle_q^2} .
\end{equation}
With the  definition of symmetry-resolved spread complexity used in
\cite{Caputa:2025mii, Caputa:2025ozd}, fixed charge alone does \emph{not} turn $t^2$ into $t$.
\\
A plausible linear term appears only if one asks a different, ``hybrid question''.
\\
Construct the \emph{total} Krylov basis $|K_n\rangle$ from the full state $|\psi\rangle$ and define
\begin{equation}
 \mathcal N=\sum_n n|K_n\rangle\langle K_n|,
 \qquad
 \widetilde C_q(t)=\langle\psi_q(t)|\mathcal N|\psi_q(t)\rangle .
\end{equation}
This is not the sector-adapted complexity; it measures the fixed-charge component in the total Krylov frame.  The key point is that $\Pi_q|\psi\rangle$ is generically not localized at $n=0$ in this basis.  For example,
\begin{equation}
 |K_0\rangle=|\psi\rangle,
\qquad
 |K_1\rangle=\frac{(H-a_0)|\psi\rangle}{b_1},
 \qquad
 \langle K_1|\psi_q\rangle
 =\frac{\sqrt{p_q}}{b_1}\big(\langle H\rangle_q-a_0\big),
\end{equation}
which is generically non-zero.  Hence $\dot{\widetilde{{C}}}_q(0)$ is generally non-zero (in correspondence with $P_y(t=0)\neq 0$ in holography), and the complexity can have an expansion
\begin{equation}
 \widetilde C_q(t)=J_q\,t+O(t^2),\qquad
 J_q=i\langle\psi_q|[H,\mathcal N]|\psi_q\rangle .
\end{equation}
%In the total Krylov basis, where
%$H=\sum_n a_n|K_n\rangle\langle K_n|+\sum_n b_{n+1}(|K_{n+1}\rangle\langle K_n|+\mathrm{h.c.})$, this current is
%\begin{equation}
% J_q=2\sum_n b_{n+1}\,\mathrm{Im}\!\big(v_{n+1}^{(q)*}v_n^{(q)}\big),
% \qquad v_n^{(q)}=\langle K_n|\psi_q\rangle .
%\end{equation}
Therefore a genuine short-time linear term requires a non-zero initial Krylov current $J_q$.  Symmetry resolution can make such a current possible because the charge projection delocalises the state in the total Krylov chain and because the total Krylov number operator $\mathcal N$ need not decompose simply into charge-sector Krylov number operators.
\\
This is a plausibility argument, to understand that systems with conserved charges might lead to different behaviours of the complexity depending on the definition of the projection on a given basis.

\section{About the definition of momentum and proper momentum}\label{appendixBproper}
In this appendix, we aim to review the geodesic equations obtained from a Lagrangian formalism in more detail \cite{Hobson:2006se}. Then, we will make some comments on the momentum of the falling particle along the geodesic and the definition of the proper momentum used in this paper. 

We describe the geodesic curve $x^a(u)$ in terms of some general (not necessarily affine) parameter $u$. The length along the curve is

\begin{eqnarray}
L=\int_A^B d s=\int_A^B\left|g_{a b} \dot{x}^a \dot{x}^b\right|^{1 / 2} d u,   
\end{eqnarray}
 with A and B being the initial and final points along the curve, and the overdot representing $d / d u$. The Euler-Lagrange equations read

\begin{eqnarray}
\frac{d}{d u}\left(\frac{g_{a c} \dot{x}^a}{\dot{s}}\right)-\frac{1}{2 \dot{s}}\left(\partial_c g_{a b}\right) \dot{x}^a \dot{x}^b=0 .
\end{eqnarray}

Noting that $\dot{g}_{a c}=\left(\partial_b g_{a c}\right) \dot{x}^b$, the LHS can be written as 

\begin{eqnarray}
\frac{d}{d u}\left(\frac{g_{a c} \dot{x}^a}{\dot{s}}\right)=\frac{1}{\dot{s}}\left[\left(\partial_b g_{a c}\right) \dot{x}^a \dot{x}^b+g_{a c} \ddot{x}^a-\frac{\ddot{s}}{\dot{s}} g_{a c} \dot{x}^a\right] .
\end{eqnarray}
Which, after substituting back into the Euler-Lagrange equations, gives
\begin{eqnarray}
g_{a c} \ddot{x}^a+\left(\partial_b g_{a c}\right) \dot{x}^a \dot{x}^b-\frac{1}{2}\left(\partial_c g_{a b}\right) \dot{x}^a \dot{x}^b=\left(\frac{\ddot{s}}{\dot{s}}\right) g_{a c} \dot{x}^a .
\end{eqnarray}
One can use $\left(\partial_b g_{a c}\right) \dot{x}^a \dot{x}^b=\frac{1}{2}\left(\partial_b g_{a c}+\right. \left.\partial_a g_{b c}\right) \dot{x}^a \dot{x}^b$. in the previous equation and multiply the whole equation by $g^{d c}$ to find
\begin{eqnarray}
\ddot{x}^d+\frac{1}{2} g^{d c}\left(\partial_b g_{a c}+\partial_a g_{b c}-\partial_c g_{a b}\right) \dot{x}^a \dot{x}^b=\left(\frac{\ddot{s}}{\dot{s}}\right) \dot{x}^d .
\end{eqnarray}

Making use of the expression for the connection in terms of the metric and relabeling some of the indices, one has
\begin{eqnarray}
\ddot{x}^a+\Gamma^a{ }_{b c} \dot{x}^b \dot{x}^c=\left(\frac{\ddot{s}}{\dot{s}}\right) \dot{x}^a .
\end{eqnarray}

 In affine coordinates, the RHS should vanish. Hence, for a non-null geodesic, we see that an affine parameter $u$ can be obtained from the distance $s$ measured along the geodesic by the linear relation $u=ps+q$. Here, $p$ and $q$ are constants and $(p \neq 0)$.

As a side note, if one starts with the auxiliary Lagrangian
\begin{eqnarray} \label{eq:Lgeo}
L=g_{a b} \dot{x}^a \dot{x}^b , 
\end{eqnarray}
instead, the Euler-Lagrange equations will simply read
\begin{eqnarray}
\ddot{x}^a+\Gamma^a{ }_{b c} \dot{x}^b \dot{x}^c=0,
\end{eqnarray}
which is the geodesic equation in the affine parameter. Nowhere in the process, the $\dot s=0$ is required, so it is applicable to both null and non-null geodesics.
%%%%
\subsection{Momentum calculation}
I this section, we calculate the momentum of a particle freely falling in confining backgrounds of the form 
\begin{equation}
    ds_4^2 =\frac{r^2}{l^2}\left( -dt^2+dx^2+ f(r) d\phi^2\right)+ \frac{l^2 dr^2}{r^2 f(r)}\equiv\frac{l^2}{z^2}\left( -dt^2+dx^2+ f(z) d\phi^2\right)+ \frac{l^2 dz^2}{z^2 f(z)}
\end{equation}
and try to derive some general results.

We aim to calculate the momentum four vector, $\mathbf{p}$, defined in coordinate basis as $p^\mu=\frac{d x^\mu}{d\tau}$, $\tau$ being the proper time measured by the in-falling particle, which is an invariant quantity under Lorentz transformations. 

The equations of motion for the freely falling particle, with trajectory parametrized by an affine parameter $\tau$, are given by
\begin{equation} \label{eq:geo}
    \ddot{x}^a+\Gamma^a_{bc} \dot{x}^b \dot{x}^c=0,
\end{equation}
where the derivative is with respect to $\tau$.
 For the background of the form
 \begin{equation}
     ds^2=-\frac{1}{z^2}dt^2+\frac{1}{z^2f(z)}dz^2+\cdots,
 \end{equation}
using eq. \eqref{eq:geo} or eq. \eqref{eq:Lgeo} one has,
\begin{eqnarray} \label{eq:t}
    &\frac{d}{d\tau}(\frac{\dot t}{z^2})=0 \quad \Rightarrow \quad \frac{\dot t}{z^2}=k, \\ \nonumber
    &\frac{d}{d\tau}(\frac{-2\dot z}{z^2f(z)})=\frac{\partial}{\partial z}(\frac{\dot t^2}{z^2}-\frac{\dot z^2}{z^2f(z)}),
\end{eqnarray}
and the constraint from the norm of the timelike momentum vector,
\begin{equation}
    p^\mu p_\mu=\frac{\dot t^2}{z^2}-\frac{\dot z^2}{z^2f(z)}=1.
\end{equation}
 An observer at infinity will have a four-velocity $\mathbf{u}=(-1,0,0,0)$ and it measures the energy of the particle at $z\rightarrow0$ as 
 \begin{eqnarray}
     \label{eq:E} E=\mathbf{p}\,\cdot\mathbf{u}=p_0\,u^0=p_0=g_{00}p^0=g_{00}\dot t=k.
 \end{eqnarray}
 Hence, we assign $k\equiv H$. By inserting eq.\eqref{eq:E} in eq.\eqref{eq:t} one has 
 \begin{eqnarray}
     &\frac{dt}{d\tau}&=H z^2,\,\\
     &\frac{dz}{d\tau}&=z\sqrt{(H^2z^2-1)f(z)},
 \end{eqnarray}
 hence $\mathbf{p}=(H z^2,z\sqrt{(H^2z^2-1)f(z)},0,0)$. We can obtain the relations
 \begin{eqnarray}
     \frac{dz}{dt}=\frac{\sqrt{(H^2z^2-1)f(z)}}{H z}.
 \end{eqnarray}
 and
 \begin{eqnarray}
     p_z=g_{zz}p^z=\frac{\sqrt{(H^2z^2-1)}}{z\sqrt{f(z)}}.
 \end{eqnarray}
 matching with previous results.
 While the proper momentum reads as
 \begin{eqnarray}
     p_y\equiv\frac{1}{\sqrt{g_{zz}}}p_z =\sqrt{(H^2z^2-1)}.
 \end{eqnarray}
 Note that the function $f(z)$, which has a root in $z=z^*$ has disappeared from the momentum expression, hence the proper momentum is always finite along the fall. It is noteworthy that while $\dot{z}$ might vanish at the end of space or $p_z$ might have poles at that point, $p_y$ will always remain finite.
 
\section{Complexity for a QFT coupled to gravity}\label{lastapp}
In all the examples dealt with in this work, we have considered the holographic dual of a QFT in diverse dimensions, displaying confinement. The QFT is {\it decoupled} from gravity. 

Here, we consider a situation in which the QFT in the far IR is ${\cal N}=4$ SYM deformed by a dimension eight operator \cite{Intriligator:1999ai}. The needed UV completion this system has is in terms of coupling the QFT to ten-dimensional gravity (IIB strings). The holographic dual metric to this QFT in this case is that of D3 branes before taking the decoupling limit of Maldacena (near horizon limit).
\begin{equation}
    ds_E^2=\hat{h}^{-\frac{1}{2}}\Big[-dt^2+ d\vec{x}_3^2 +\hat{h}(dr^2+r^2 d\Omega_5) \Big].~~~F_5\sim(1+*)d\hat{h}^{-1}.\label{d3out}
\end{equation}
As above, we do not need the details of the RR five form. The relevant functions to apply the treatment in Section \ref{sec:3.1} are,
\begin{equation}
 A(r)=\hat{h}^{-\frac{1}{2}}(r), ~~B(r)=\hat{h}(r),~~\hat{h}(r)=\mu+\frac{Q^4}{r^4}, ~Q^4= 2\pi g_s\alpha'^2 N.\label{fucntionsd3out}  
\end{equation}
The parameter $\mu$ is related to the coupling of the dimension eight operator $O_8\sim \mu F_{\mu\nu}^4$.

\begin{figure}
    \centering
    \includegraphics[width=0.5\linewidth]{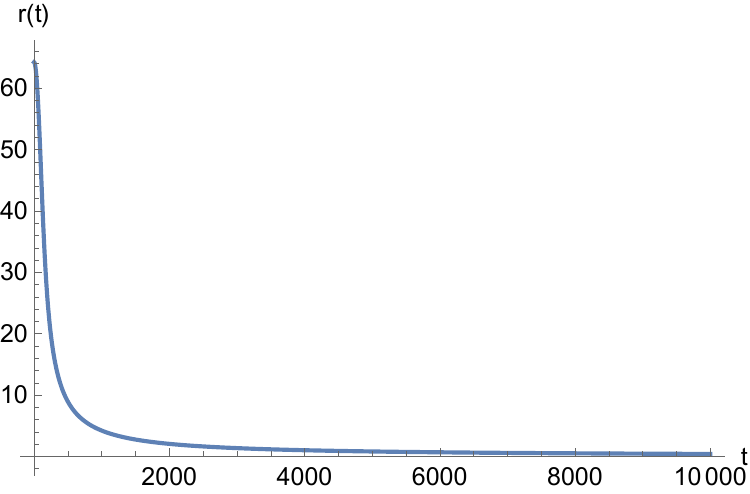}
        \includegraphics[width=0.5\linewidth]{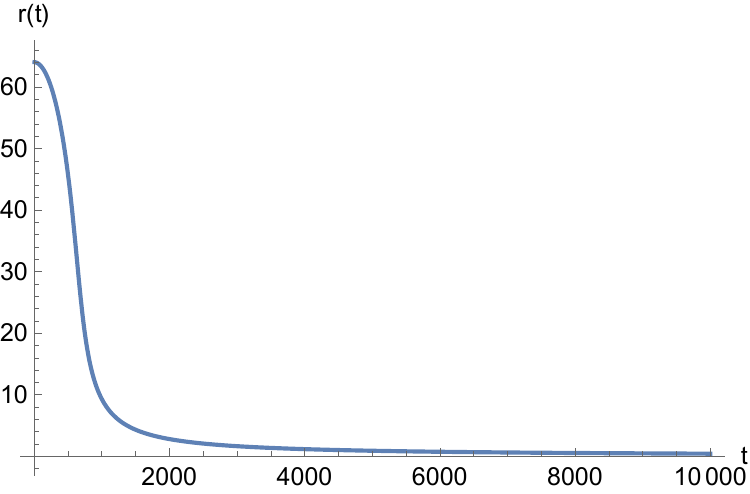}
    \caption{$r(t)$  with $N_c=10$ and  $\mu = 1, H=0.85, m=1$ (upper panel) while $\mu = 10, H=0.55, m=1$ (lower panel)}
    \label{fig:rtmu}
\end{figure}

The action, equation of motion and complexity calculated using a massive geodesic that radially falls in this background from a position $r_{UV}$ and with zero initial velocity are,
\begin{eqnarray}
& & S=-m \int dt \sqrt{\hat{h}^{-\frac{1}{2}}(r)\left( 1-\hat{h}(r) \dot{r}^2\right) }, \label{d3outeqs}\\
& & \pm\frac{(t-t_0)}{H}=\int dr \frac{\hat{h}^{\frac{3}{4}}(r)}{\sqrt{H^2 \hat{h}^{\frac{1}{2}} (r) - m^2}}, ~~~~\dot{C}(t)\sim m\sqrt{\left( \frac{\hat{h}(r)}{\hat{h}(r_{UV})}\right)^{\frac{1}{2}}-1}.\nonumber
\end{eqnarray}
For $\mu\simeq 0$ the expressions are well approximated by the conformal case, see eqs.(\ref{resultT11}). Figure \ref{fig:rtmu}.
shows $r(t)$ for different values of $(\mu,H)$. The particle falls more slowly when it is positioned in a place with less curvature; hence, the complexity grows initially more slowly in these cases. 

\bibliographystyle{JHEP}
\bibliography{main.bib}
\end{document}